\documentclass[format=acmsmall, review=false]{acmart}

\usepackage{acm20} 
\usepackage{graphicx} 
\usepackage{algorithm}
\usepackage[noend]{algpseudocode}

\title{Snowman for partial synchrony}

\author{Aaron Buchwald}
\affiliation{%
\institution{ \institution{Ava Labs}   \city{NY} \country{USA}}
}

\author{Stephen Buttolph}
\affiliation{%
\institution{ \institution{Ava Labs}  \city{NY} \country{USA}}
}

\author{Andrew Lewis-Pye}
\affiliation{%
\institution{ \institution{London School of Economics}  \city{London} \country{UK}}
}

\author{Kevin Sekniqi}
\affiliation{%
\institution{ \institution{Ava Labs}  \city{SF} \country{USA}}
}

\date{March 2025}

\begin{document}

\begin{abstract}
    Snowman is the consensus protocol run by blockchains on Avalanche.  A recent paper \cite{buchwald2024frosty} established a rigorous proof of probabilistic consistency for Snowman in the \emph{synchronous} setting, under the simplifying assumption that correct processes execute sampling rounds in `lockstep'. In this paper, we describe a modification of the protocol that ensures consistency in the \emph{partially synchronous} setting, and when correct processes carry out successive sampling rounds at their own speed, with the time between sampling rounds determined by local message delays.  
\end{abstract}

\maketitle


\section{Introduction} \label{intro}

While recent years have seen significant interest in applications of blockchain technology, the \emph{scalability} challenge has provided a barrier to large-scale adoption: how can a blockchain maintain low latency and high throughput when the number of participants $n$ is large? As a consequence, the last decade has produced a substantial amount of research \cite{yin2019hotstuff,spiegelman2022bullshark,shrestha2024sailfish,lewis2024lumiere,naor2024expected,bagaria2019prism} aimed at meeting this challenge. The \emph{communcation cost} of consensus is one of the fundamental considerations in this regard, since the lower bound of Dolev and Reischuk \cite{dolev1985bounds} asserts that deterministic protocols require $\Omega(n^2)$ communication complexity per consensus decision: deterministic protocols that can tolerate a Byzantine (i.e. arbitrary) adversary that controls $O(n)$ participants must necessarily suffer a quadratic blow-up in communication cost as the size of the network grows.

\vspace{0.3cm} 
 The Snow family of probabilistic consensus protocols was introduced in the Avalanche whitepaper \cite{rocket2019scalable} as a novel approach to the scalability challenge. Today, Avalanche implements a member of this family called Snowman. A major advantage of Snowman is that each consensus decision only requires an expected constant communication overhead per participant in the `common' case that the protocol is not under substantial Byzantine attack, i.e.\ it provides a probabilistic solution to the scalability problem which ensures that the expected communication overhead per participant is independent of the total number of participants $n$ during normal operation. On the other hand, the two following concerns had remained: 
    \begin{enumerate}
        \item The probabilistic nature of the sampling procedures used by Snowman mean that it is complex to analyze. 
        \item Liveness attacks exist in the case that a Byzantine adversary controls more than $O(\sqrt{n})$ participants, slowing termination to more than a logarithmic number of steps.
    \end{enumerate}
    
\vspace{0.3cm} 
A recent paper \cite{buchwald2024frosty} addressed these issues by establishing a rigorous proof of probabilistic consistency for Snowman in the \emph{synchronous} setting (where communication is always reliable), and under the simplifying assumption that correct participants execute sampling rounds in `lockstep'. That paper also described how to augment Snowman with a  \emph{liveness module}, named Frosty, that can be triggered in the case that a substantial adversary launches a liveness attack, and which guarantees liveness in this event by temporarily forgoing the communication complexity advantages of Snowman, but without sacrificing these low communication complexity advantages during normal operation. While the proof of consistency for Snowman assumes synchrony, the liveness module utilizes a classical `quorum-based' protocol that functions in partial synchrony, i.e.\ even if communication is sometimes unreliable.

\vspace{0.3cm} 
\noindent \textbf{Modifying Snowman for partial synchrony}. Two weaknesses of the analysis in \cite{buchwald2024frosty} are: 
    \begin{enumerate}
        \item The consistency analysis assumes synchrony. While protocols such as Bitcoin require synchrony to guarantee consistency, quorum based protocols often offer the much stronger guarantee that consistency holds even when communication is not reliable. Ideally, one would like a version of Snowman that gives such guarantees. 
        \item The analysis makes the simplifying assumption that correct participants execute sampling rounds in `lockstep'.
    \end{enumerate}
To address these issues, this paper  describes a modification of Snowman, called Snowman$^{\diamond}$, which ensures probabilistic consistency even when communication is not reliable (it functions in \emph{partial synchrony}) and without the assumption that correct participants execute sampling rounds in lockstep. In Snowman$^{\diamond}$, correct participants may carry out successive sampling rounds at their own speed, with the time between sampling rounds determined by local message delays.
Since the Frosty module \cite{buchwald2024frosty} can be used to ensure liveness, we focus on the analysis of probabilistic consistency for Snowman$^{\diamond}$ in this paper.

\vspace{0.2cm} 
\noindent \textbf{The structure of this paper}.  Section \ref{model} describes our formal model, which is standard for papers in distributed computing. To describe and analyze Snowman$^{\diamond}$, which is a protocol for State Machine Replication, we first describe and analyze Snowflake$^{\diamond}$, which is a simpler protocol for Binary Byzantine Agreement. To motivate our specification of Snowflake$^{\diamond}$, in Section \ref{recallingplus}, we first recall Snowflake$^+$ (first described in \cite{buchwald2024frosty}), which is a protocol for Binary Byzantine Agreement in the synchronous setting. We also recall the analysis of Snowflake$^+$ from \cite{buchwald2024frosty}  (which assumes the synchronous setting) in Section \ref{Snowflakeanalysis}. Section \ref{changes} gives some intuition behind the modifications required to transform Snowflake$^+$ into a protocol (Snowflake$^{\diamond}$) that functions in partial synchrony. 
 Section \ref{SnowFD} then gives the formal specification of Snowflake$^{\diamond}$, while 
 section \ref{SnowFDanal} formally establishes probabilistic agreement for Snowflake$^{\diamond}$ in the partially synchronous setting. Section \ref{SnowmanD} gives the formal specification of Snowman$^{\diamond}$ (a protocol for State Machine Replication, rather than Binary Byzantine Agreement).  
Section \ref{consisanal} then formally establishes probabilistic consistency for Snowman$^{\diamond}$ in the partially synchronous setting.  Section \ref{qf} describes conditions for quick finality.  Finally, Section \ref{rw} describes related work.


\section{The model} \label{model}

We consider a set $\Pi= \{ p_0,\dots, p_{n-1} \}$ of $n$ processes. Process $p_i$ is told $i$ as part of its input. For the sake of simplicity, we assume a static adversary that controls up to $f$ processes, where $f$ is a known bound. Generally, we will assume $f<n/5$. The bound $f<n/5$ is chosen only so as to make our consistency proofs as simple as possible, and providing an analysis for larger $f$ is the subject of future work. 
 A process that is controlled by the adversary is referred to as \emph{Byzantine}, while processes that are not Byzantine are \emph{correct}. Byzantine processes may display arbitrary behaviour, modulo our cryptographic assumptions (described below).

\vspace{0.2cm} 
\noindent \textbf{Cryptographic assumptions}. Our cryptographic assumptions are standard for papers in distributed computing. Processes communicate by point-to-point authenticated channels. We use a cryptographic signature scheme, a public key infrastructure (PKI) to validate signatures, and a collision-resistant hash function $H$. 
 We assume a computationally bounded adversary. Following a common standard in distributed computing and for simplicity of presentation (to avoid the analysis of certain negligible error probabilities), we assume these cryptographic schemes are perfect, i.e.\ we restrict attention to executions in which the adversary is unable to break these cryptographic schemes. In a given execution of the protocol, hash values are therefore assumed to be unique.

 \vspace{0.2cm} 
\noindent \textbf{Communication and clocks}. As noted above, processes communicate using point-to-point authenticated channels. We consider the standard synchronous and partially synchronous settings:

\vspace{0.1cm} 
\noindent \emph{The synchronous setting}. For some known $\Delta$, a message sent at time $t$ must arrive by time $t + \Delta$.

\vspace{0.1cm} 
\noindent \emph{The partially synchronous setting}. For some known $\Delta$ and some unknown \emph{Global-Stabilization-Time} (GST), a message sent at time $t$ must arrive by time $\text{max} \{ t, \text{GST} \} +\Delta$. This setting formalizes the requirement that consistency must be maintained in asynchrony (when there is no bound on message delays), but liveness need only be ensured given sufficiently long intervals of synchrony. 

\vspace{0.1cm}
For the new results of this paper, we do not suppose that the clocks of correct processes are synchronized. For the sake of simplicity, however, we do suppose that the clocks of correct processes all proceed in real time, i.e.\ if $t'>t$ then the local clock of correct $p$ at time $t'$ is $t'-t$ in advance of its value at time $t$. This assumption is made only for the sake of simplicity, and our arguments are easily adapted to deal with a setting in which there is a given upper bound on the difference between the clock speeds of correct processes. 
 Later, in Section \ref{qf}, we will also consider a setting in which there is a known bound $\Delta^*$ between the clocks of correct processes. 


 \vspace{0.2cm} 
\noindent \textbf{The binomial distribution}. Consider $k$ independent and identically distributed random variables, each of which has probability $x$ of taking the value `red'. We let $\text{Bin}(k,x,m)$ denote the probability that $m$ of the $k$ values are red, and we write $\text{Bin}(k,x,\geq m)$ to denote the probability that \emph{at least} $m$ values are red (and similarly for $\text{Bin}(k,x,\leq m)$).

 \vspace{0.2cm} 
\noindent \textbf{Dealing with small probabilities}. In analysing the security of a cryptographic protocol, one standardly regards a function $f:\mathbb{N} \rightarrow \mathbb{N}$ as \emph{negligible} if, for every $c\in \mathbb{N}$, there exists $N_c\in \mathbb{N}$ such that, for all $x\geq N_c$, $|f(x)|<1/x^c$. Our concerns here, however, are somewhat different. As noted above, we assume the cryptographic schemes utilized by our protocols are perfect. For certain \emph{fixed} parameter values (e.g.\ setting $n=250$, $k=80$, $\alpha_1=41$, $\alpha_2=72$ and $\beta=12$ in an instance of Snowflake$^+$, as described in Section \ref{recallingplus}), we want to be able to argue that error probabilities are sufficiently small that they can reasonably be dismissed.

In our analysis, we will therefore identify certain events as occurring with \emph{small} probability (e.g. with probability $<10^{-20}$), and may then condition on those events not occurring. 
Often, we will consider specific events, such as the probability in a round-based protocol that a given process performs a certain action $x$ in a given round. In dismissing small error probabilities, one then has to take account of the fact that there may be many opportunities for an event of a given type to occur, e.g. any given process may perform action $x$ in any given round. How reasonable it is to condition on no correct process performing action $x$ may therefore depend on the number of processes and the number of rounds, and we assume `reasonable' bounds on these values. We will address such accountancy issues as they arise.

\section{Recalling Snowflake$^{+}$.} \label{recallingplus}

\textbf{The analysis of \cite{buchwald2024frosty}: Snowman via Snowflake$^+$ in the lockstep model}. Towards providing a rigorous analysis for Snowman in the synchronous setting, the route taken in \cite{buchwald2024frosty} was to assume a `lockstep' model, in which (the clocks of correct processes are synchronized and) correct processes are required to execute the instructions for each round of the protocol at the same time as each other. Rather than jumping straight to an analysis of the Snowman protocol, which is designed to carry out state-machine-replication (SMR), that paper first described a simple probabilistic protocol for Binary Byzantine Agreement, called Snowflake$^+$. It was then shown how to adapt Snowflake$^+$ to give the Snowman protocol for SMR, and how to adapt the security analysis of Snowflake$^+$ to establish a rigorous proof of consistency for Snowman in the synchronous lockstep model. 

\vspace{0.2cm} 
\noindent \textbf{Moving to partial synchrony: Snowman$^{\diamond}$ via Snowflake$^{\diamond}$}. In this paper, our aim is to modify Snowman so as to ensure consistency in the partially synchronous setting, and to permit each process to progress through the protocol instructions as fast as local message delivery will allow. To present our modified version of Snowman, however, we adopt the same basic structure as in \cite{buchwald2024frosty}: we first present a simple protocol for Binary Byzantine Agreement (that now operates effectively outside the lockstep model and in partial synchrony) and then show how to adapt that protocol to give a protocol for SMR (that also now operates effectively outside the lockstep model and in partial synchrony). Our new probabilistic protocol for Binary Byzantine Agreement is called Snowflake$^{\diamond}$.  We will refer to the new protocol for SMR that builds on Snowflake$^{\diamond}$ as Snowman$^{\diamond}$. 

\vspace{0.2cm} 
 To motivate the definition of our new protocol for Binary Byzantine Agreement, Snowflake$^{\diamond}$, in this section we first recall the  description and analysis of Snowflake$^+$ (assuming synchrony) from \cite{buchwald2024frosty}. In Section \ref{changes} we discuss changes to Snowflake$^+$ that are required to drop the lockstep model and deal with partial synchrony.  Snowflake$^{\diamond}$  is then defined precisely in Section \ref{SnowFD}.  

\vspace{0.2cm} 
 \noindent \textbf{Binary Byzantine Agreement}.
 First, let us recall the requirements for Binary Byzantine Agreement.  Each process $p_i$ begins with a value $\mathtt{input}_i\in \{ 0, 1 \}$.   A probabilistic protocol for Binary Byzantine Agreement is required to satisfy the following properties, except with small error probability: 

 \noindent \emph{Agreement}: No two correct processes output different values.   

 \noindent \emph{Validity}: If every correct process $p_i$ has the same value $\mathtt{input}_i$, then no correct process outputs a value different than this common input.

 \noindent \emph{Termination}: Every correct process gives an output.



\vspace{0.2cm} 
\noindent \textbf{Recalling Snowflake$^+$}. The protocol parameters are  $k,\alpha_1,\alpha_2,\beta \in \mathbb{N}_{>0}$ and satisfy the constraints that $\alpha_1>k/2$ and $\alpha_2\geq \alpha_1$. Each process $p_i$ maintains a variable $\mathtt{val}_i$, initially set to $\mathtt{input}_i$. The parameter $k$ determines sample sizes. The parameter $\alpha_1$ is used to determine when  $p_i$ changes $\mathtt{val}_i$. Parameters $\alpha_2$ and $\beta$ are used to determine the conditions under which $p_i$ will output and terminate.   

\vspace{0.2cm}
\noindent \textbf{The protocol instructions for Snowflake$^+$}. All correct processes begin the protocol (with clocks synchronized) at time 0. The instructions are divided into rounds, with round $s$ occurring at time $2\Delta s$.  In round $s$, process $p_i$:
\begin{enumerate} 
\item Sets $\langle p_{1,s},\dots p_{k,s} \rangle$ to be a sequence of $k$ processes (specific to $p_i$). For $j\in [1,k]$,  $p_{j,s}$ is sampled from the uniform distribution\footnote{In proof-of-stake implementations, sampling will be stake-weighted. For the sake of simplicity of presentation, we ignore such issues here.} on all processes (so sampling is ``with replacement''). 
\item Requests each $p_{j,s}$ (for $j\in [1,k]$) to report its present value $\mathtt{val}_j$.  
\item Waits time $\Delta$ and reports its present value $\mathtt{val}_i$ to any process that has requested it in round $s$.
\item Waits another $\Delta$ and considers the values reported in round $s$:
\begin{itemize}
 \item If at least $\alpha_1$ of the reported values are $1-\mathtt{val}_i$, then $p_i$ sets $\mathtt{val}_i:= 1-\mathtt{val}_i$.
 \item If $p_i$ has seen $\beta$ consecutive rounds in which at least $\alpha_2$ of the reported values are equal to $\mathtt{val}_i$, then $p_i$ outputs this value and terminates.
\end{itemize}

\end{enumerate}

The pseudocode is described in Algorithm 1.


\begin{algorithm} \label{pc:Snowflake}
\caption{Snowflake$^{+}$: The instructions for  $p_i$}
\begin{algorithmic}[1]

    \State \textbf{Inputs} 
    \State $\mathtt{input}_i\in \{ 0, 1 \}$              \Comment $p_i$'s input
    \State $\Delta, k,\alpha_1,\alpha_2,\beta \in \mathbb{N}$  \Comment Protocol parameters

    \State \textbf{Local variables} 
    
   \State $\mathtt{val}_i$, initially set to $\mathtt{input}_i$   \Comment $p_i$'s present `value'      
    \State $\mathtt{count}$, initially set to $0$ \Comment Output once $\mathtt{count}$ reaches $\beta$
    \State $v_i(j,s)$, initially undefined           \Comment Stores at most one received value per round
     
    \State 

  \State \textbf{The instructions for round $s$, beginning at time $2\Delta s$:}

  \State \hspace{0.3cm} Form sample sequence $\langle p_{1,s},\dots p_{k,s} \rangle$;           \Comment Sample  with replacement

  \State \hspace{0.3cm} For $j\in [1,k]$, send $s$ to $p_{j,s}$;    \Comment Ask  $p_{j,s}$ for present value

   \State \hspace{0.3cm}  Wait $\Delta$;

   \State \hspace{0.3cm} For each $j$ such that $p_i$ has received $s$ from $p_j$:

   \State \hspace{0.8cm} Send $(s,\mathtt{val}_i)$ to $p_j$;

   \State \hspace{0.3cm}  Wait $\Delta$;

   \State \hspace{0.3cm} For each $j\in [1,k]$:
x   \State \hspace{0.8cm} \textbf{If} $p_i$ has received a first message $(s,v)$ from $p_{j,s}$;
   \State \hspace{1.3cm}  Set $v_i(j,s):=v$;
   \State \hspace{0.8cm} \textbf{Else} set $v_i(j,s):= \bot$;

   \State \hspace{0.3cm} \textbf{If} $|\{ j: 1\leq j \leq k, v_i(j,s)==1-\mathtt{val}_i \}|\geq \alpha_1$, set $\mathtt{val}_i:=1-\mathtt{val}_i$, $\mathtt{count}:=0$;

  \State \hspace{0.3cm} \textbf{If} $|\{ j: 1\leq j \leq k, v_i(j,s)==\mathtt{val}_i \}|< \alpha_2$, set $\mathtt{count}:=0$;

  \State \hspace{0.3cm} \textbf{If} $|\{ j: 1\leq j \leq k, v_i(j,s)==\mathtt{val}_i \}| \geq  \alpha_2$, set $\mathtt{count}:=\mathtt{count}+1$;

  \State \hspace{0.3cm} \textbf{If} $\mathtt{count}\geq \beta$, output $\mathtt{val}_i$ and terminate.

\end{algorithmic}
\end{algorithm}



\subsection{Analyzing Snowflake$^+$}  \label{Snowflakeanalysis}

We assume $f<n/5$ and consider the synchronous setting. For the sake of concreteness, we fix a specific set of parameter values.\footnote{As for all arguments presented in the paper, the following analysis of Snowflake$^+$ is easily adapted to deal with alternative parameter values. If we fix $\alpha_1:=\lfloor k/2 \rfloor +1$, then error probabilities will be smaller for larger values of $\alpha_2$ and $\beta$. For smaller values of $\alpha_2$, similar error probabilities can be obtained by increasing $\beta$ -- the required values for $\beta$ are easily found by adapting the corresponding binomial calculations.} To motivate the definition of Snowflake$^{\diamond}$, we establish satisfaction of agreement (no two correct processes output different values) for $k=80$, $\alpha_1=41$, $\alpha_2=72$, and $\beta=12$, under the assumption that the population size $n\geq 250$. Ultimately, we will be able to give a proof of agreement for Snowflake$^{\diamond}$ which is similar to that for Snowflake$^+$ but which does not assume synchrony, and will then be able to adapt that proof to give a proof that Snowman$^{\diamond}$ satisfies consistency in the partially synchronous setting. In this paper, we are focused on establishing consistency for Snowman$^\diamond$. We therefore do not address termination or validity for Snowflake$^+$ or Snowflake$^{\diamond}$.\footnote{For an analysis of non-trivial conditions under which termination is satisfied except with small error probability, see \cite{amores2024analysis}: the analysis there is stated in terms of the Slush protocol, but the conclusion that the protocol reaches a stable consensus in $O(\text{log }n)$ rounds, and that this holds even when the adversary can influence up to $O(\sqrt{n})$ processes, carries over directly to the protocols considered in this paper.} 

In analyzing Snowflake$^+$ (as well as Snowflake$^\diamond$ and Snowman$^\diamond$), we make the assumption that $f<n/5$ and $n\geq 250$ only so as to be able to give as simple a proof as possible: a more fine-grained analysis for smaller $n$ is the subject of future work.

\vspace{0.2cm}
\noindent \textbf{Coloring the processes}. Since 0 and 1 are not generally used as adjectives, let us say a correct process $p_i$ is `blue' in round $s$ if $\mathtt{val}_i=0$ at the beginning of round $s$, and that $p_i$ is `red' in round $s$ if $\mathtt{val}_i=1$ at the beginning of round $s$. Recall (from Algorithm 1) that $v_i(j,s)$ is the color that $p_{j,s}$ (as specified by $p_i$'s sample) reports to $p_i$ in round $s$.  We'll say a correct process $p_i$ `samples $x$ blue' in round $s$ if $|\{ j: 1\leq j \leq k, v_i(j,s)=0 \}|=x$ (and similarly for red).  We'll also extend this terminology in the obvious way, by saying that a process outputs `blue' if it outputs 0 and outputs `red' if it outputs 1. In the below, we'll focus on the case that, in the first round in which a correct process outputs (should such a round exist), some correct process outputs red. A symmetric argument can be made for blue.

\vspace{0.2cm} In the following argument, we will adopt the conventions described in Section \ref{model} concerning the treatment of small error probabilities. We will identify certain events as occurring with \emph{small} probability, and may then condition on those events not occurring. Where there are multiple opportunities for an event of a certain type to occur, one must be careful to account for the accumulation of small error probabilities. To deal with the accumulation of small error probabilities, we suppose that at most 10,000 processes execute the protocol for at most 1000 years, executing at most 5 rounds per second. 

\vspace{0.2cm}
\noindent \textbf{Establishing Agreement}. The argument consists of three parts: 

\vspace{0.2cm} 
\noindent \textbf{Part 1}. First, let us consider what happens when the proportion of correct processes that are red reaches a certain threshold. In particular, let us consider what happens when at least 75\% of the correct  processes are red in a given round $s$. A direct calculation for the binomial distribution shows that the probability a given correct process is red in round $s+1$ is then at least 0.9555, i.e.\ $\text{Bin}(80,0.8 \times 0.75,\geq41)>0.9555$. Assuming a population of at least 250, of which at least 80\%  are correct (meaning that at least 200 are correct), another direct calculation for the binomial distribution shows that the probability that it fails to be the case that at least  75\% of the correct processes are red in round $s+1$ is upper bounded by $4\times 10^{-24}$, i.e.\ $\text{Bin}(n,0.9555,< 3n/4)<4 \times 10^{-24}$ for $n\geq 200$. 
Note that this argument requires no knowledge as to the state of each process in round $s$, other than the fact that at least 75\% of the correct  processes are red. 

The analysis above applies to any single given round $s$. Next, we wish to iterate the argument over rounds in order to bound the probability that the following statement holds for \emph{all} rounds: 
\begin{enumerate} 
\item[$(\dagger_1)$] If at least 75\% of the correct processes are red in any round $s$, then, in all rounds $s'\geq s$, at least 75\% of the correct processes are red.
\end{enumerate}
\noindent To bound the probability that $(\dagger_1)$ fails to hold, we can bound the number of rounds, and then apply the union bound to our analysis of the error probability in each round. Suppose that the protocol is executed for at most 1000 years, with at most 5 rounds executed per second.  This means that less than $1.6\times 10^{11}$ rounds are executed. The union bound thus gives a cumulative error probability of less than $(1.6\times 10^{11}) \times (4\times 10^{-24})  < 7\times 10^{-13}$, meaning that $(\dagger_1)$ fails to hold  with probability at most  $7\times 10^{-13}$. 



\vspace{0.2cm} 
\noindent \textbf{Part 2}.  Another direct calculation for the binomial distribution shows that, if \emph{at most} 75\% of correct processes are red in a given round $s$, then the probability a given correct process $p$ samples 72 or more red 
in round $s$ is upper bounded by 0.0131, i.e.\ $\text{Bin}(80,(0.75\times 0.8)+0.2,\geq 72)<0.0131$. If, for some $x\geq 1$ it then holds that at most 75\% of correct processes are red in round $s+x$, then (independent of previous events), the probability $p$ samples 72 or more red in round $s+x$ is again upper bounded by 0.0131.  So, if we consider any 12 given consecutive rounds and any given correct process $p$, the probability that at most 75\% of correct processes are red in all 12 rounds and $p$ samples at least 72 red in all 12 rounds is upper bounded by  $0.0131^{12}<10^{-22}$. If at most 10,000 processes execute the protocol for at most 1000 years, executing at most 5 rounds per second, we can then apply the union bound to conclude that the following statement fails to hold with probability at most $10^{-22}\times 10000 \times (1.6 \times 10^{11})<  2\times 10^{-7}$:

\begin{enumerate}
\item[$(\dagger_2)$] If a correct process outputs red in any round of the execution,  $s+11$ say, then, for at least one round $s'\in [s,s+11]$, at least 75\% of correct processes are red in round $s'$.
\end{enumerate}

\vspace{0.2cm} 
\noindent \textbf{Part 3}. Now we put parts 1 and 2 together. From the union bound and the analysis above, we may conclude that $(\dagger_1)$ and $(\dagger_2)$ both hold, except with probability at most 
$(7\times 10^{-13})+( 2\times 10^{-7})<3\times 10^{-7}$. So, suppose that these conditions both hold. 
According to $(\dagger_2)$,  if a correct process is the (potentially joint) first to output and outputs  red after sampling in round $s+11$, at least one round $s'\in [s,s+11]$ must satisfy the condition that at least 75\% of correct processes are red in round $s'$. From $(\dagger_1)$, it follows that at least 75\% of the correct processes must be red in all subsequent rounds.  From $(\dagger_2)$ (applied to blue), it follows that no correct process ever outputs blue. This suffices to show that Agreement is satisfied, except with small  error probability. 

\vspace{0.2cm}
\noindent \textbf{Dealing with alternative parameter values}. The argument above is easily adapted to deal with alternative parameter values. If we fix $\alpha_1:=\lfloor k/2 \rfloor +1$, then, for fixed $k$,  error probabilities will be smaller for larger values of $\alpha_2$ and $\beta$. Increasing $k$, while simultaneously increasing $\alpha_2$ in proportion to $k$, gives smaller error probabilities for fixed $\beta$ and $\alpha_1:=\lfloor k/2 \rfloor +1$. Increasing $n$ in the analysis above decreases error probabilities modulo the use of the union bound. In particular, the analysis above establishes that validity and agreement are satisfied except with small error probability for all $n\in [250,10000]$ and for all $k\geq 80$, so long as  $\alpha_1:=\lfloor k/2 \rfloor +1$, $\alpha_2\geq 0.9k$ and $\beta\geq 12$.  For smaller values of $\alpha_2$, similar error probabilities can be obtained by increasing $\beta$ -- the required values for $\beta$ are easily found by adapting the binomial calculations above.





\section{Informal discussion: Moving to partial synchrony} \label{changes} 

In Section \ref{recallingplus}, we considered the synchronous setting and supposed that correct processes  proceed through rounds in `lockstep'. We now wish to move to the partially synchronous setting, to drop the assumption that correct processes have synchronized clocks, and to define a protocol that allows an individual process to progress to round $s+1$ immediately upon receiving enough information to decide their color for that round. To produce a rigorous proof of agreement for the resulting protocol, the basic idea is that we wish to form analogues of $(\dagger_1)$ and $(\dagger_2)$ from Section \ref{recallingplus}. 

\subsection{The intuition behind forming an analogue of $(\dagger_1)$}
If processes do not proceed in lockstep through rounds, then forming an analogue of $(\dagger_1)$ is complicated by the fact that, when a correct process requests values in a given round $s$, the values reported by other correct processes may not be from round $s$. If at least 75\% of correct processes are red in round $s$, this fact alone no longer suffices to lower-bound the probability a given correct process is red in round $s+1$.  
As a first approach to dealing with this complexity, let us consider what can be inferred when we replace the assumption that at least 75\% of correct processes are red in a given round with the assumption that there exists an interval $I=[t,t+2\Delta]$ such that at least 75\% of correct processes are red at every point in $I$. We might then hope to establish that something like the following holds (except with small probability): 
\begin{enumerate} 
\item[$(\dagger_1')$] If at least 75\% of correct processes satisfy the condition that they are red for the entirety of the interval $I=[t,t+2\Delta]$ then, for every $t'\geq t$, at least 75\% of correct processes are red for the entirety of the interval $[t,t']$. 
\end{enumerate} 

\vspace{0.2cm} 
\noindent \textbf{A problem}. If processes proceed through rounds at their own speed (as fast as local message delays allow), and if message delivery times (subject to the $\Delta$ constraint after GST) are chosen by the adversary,  $(\dagger_1')$ need not hold (and, actually, this is even true in synchrony). The problem is that, subsequent to $I$, the adversary can ensure that red processes complete multiple rounds before any blue process receives enough sample values to complete a single round. During these rounds in which only red processes can change color, the proportion of correct processes that are red may drop sharply. 

\vspace{0.2cm} 
\noindent \textbf{The solution}. To remedy this issue, we modify the protocol by switching which of $\alpha_1$ and $\alpha_2$ is used to determine when a process should change color, depending on the scenario.  Recall that Snowflake considered a single $\alpha$ parameter, and required a process to sample $\alpha$ red (blue) values in a given round before changing their color to red (blue). Snowflake$^+$, on the other hand, considered two parameters $\alpha_1$ and $\alpha_2$, with $\alpha_1\leq \alpha_2$. In practical regimes, the idea is that $\alpha_1$ will generally be set lower than $\alpha$, while $\alpha_2$ might be set to a value at least $\alpha$. Snowflake$^+$ allowed a process to change color upon sampling $\alpha_1$ values of that color in a given round (while the higher value $\alpha_2$ was used only to determine when to output). We now stipulate that the protocol should proceed like Snowflake$^+$ (with $\alpha_1$ determining when to change color), except that, once a process samples $\alpha_2$ many red (blue)  in a given round, it then requires a subsequent round in which it samples $\alpha_2$ many blue (red) to change color. To be more precise,  we now stipulate that, when a process sees some round in which it samples $\alpha_2$ red (blue), it `locks' on red (blue). Once locked on red (blue), it will now not change to blue (red) unless it sees a sample in which at least $\alpha_2$ processes report being locked on blue (red) for at least $2\Delta$. To support this mechanism, sampled processes must now report not only their present color, but also how long they have been locked on that value. 

With this modification, consider the same parameter values as in Section \ref{Snowflakeanalysis}, and suppose now that we have an interval  $I=[t,t+2\Delta]$ such that at least 75\% of correct processes are locked on red for the entirety of the interval. Let the set of correct processes that are locked on red at every point in $I$ be $P_0$. Now we \emph{will} be able to argue inductively that  processes in $P_0$ will never switch to blue (except with small probability): Roughly, when each of those processes receives a sample subsequent to $I$, they receive values from a population in which (it inductively holds that)  at least 75\% of the correct processes are locked on red. Their chance of sampling at least 72 processes that have been locked on blue for at least $2\Delta$  (required for them to switch)  is then less than $\text{Bin}(80,0.8\times 0.75,\leq 8)<1.18 \times 10^{-20}$, which is small enough to dismiss. Note that the adversary can still  have the processes in $P_0$ carry out many successive rounds in the time it takes other processes to complete a single round. The difference now is that, in each round, the probability they switch to blue is sufficiently small that we can dismiss it.

 

\subsection{The intuition behind forming an analogue of $(\dagger_2)$}

 Next, we want to specify conditions for termination which suffice to ensure the following, except with small probability: when a correct process terminates and outputs red, there exists some previous interval of length $2\Delta$ during which at least 75\% of correct processes are locked on red. Our basic approach is similar to that from Section \ref{recallingplus}, but is complicated by the fact that the clocks of correct processes may not be synchronized.  
 
  Suppose that, at time $t$, $p$ sends out a request to report values. In this case, one should think of $p$ as looking for evidence to support the claim that at least 75\% of correct processes have been locked on red (or blue) during the entire interval $[t-2\Delta,t]$:
 \begin{itemize}
     \item If a response arrives from $q$ at time $t'>t$, reporting that $q$ has been locked on red for time $d$  then $p$ interprets this as $q$ having been locked on red for (at least) the entirety of the interval $[t'-d,t]$.
     \item  If the latter interval $[t'-d,t]$ includes $[t-2\Delta,t]$ then we say that the response \emph{supports outputting red}.
     \item If at least 72 out of 80 of the sample support outputting red, then the sample supports outputting red.
     \item If $\beta$ consecutive samples support outputting red, then $p$ terminates and outputs red. 
 \end{itemize}   The analysis of error probabilities will then be similar to that from Section \ref{recallingplus}.

\section{Snowflake$^{\diamond}$}  \label{SnowFD}

The pseudocode appears in Algorithm 2 (with local inputs and variables described first, and the main code appearing later).

\begin{algorithm} \label{pc:Snowflake}
\caption{Snowflake$^{\diamond}$: Inputs and local variables for  $p_i$}
\begin{algorithmic}[1]

    \State \textbf{Inputs} 
    \State $\mathtt{input}_i\in \{ 0, 1 \}$              \Comment $p_i$'s input
    \State $\Delta, k,\alpha_1,\alpha_2,\beta \in \mathbb{N}$  \Comment Protocol parameters

    \State \textbf{Local variables} 
    
   \State $\mathtt{val}_i$, initially set to $\mathtt{input}_i$   \Comment $p_i$'s present `value'   
      \State $\mathtt{val}_i(s)$, initially undefined   \Comment $p_i$'s value after responses for round $s$ 
    \State $v_i(j,s)$, initially undefined           \Comment Value reported by $p_{j,s}$

    \State $t_i(j,s)$, initially undefined            \Comment Time from which $p_i$ believes $p_{j,s}$ has been locked
     
    \State $\mathtt{start}(s)$, initially undefined   \Comment Time at which $p_i$ starts round $s$ (according to local clock)

\State $\mathtt{lock}$, initially 0        \Comment Indicates whether $p_i$ is locked on its present value

\State $\mathtt{locktime}$, initially undefined   \Comment The time at which $p_i$ most recently became locked

\State $\mathtt{lockbound}$, initially 0    \Comment Rounds after this may cause $p_i$ to become freshly locked

\State $\mathtt{newround}$, initially 1     \Comment Indicates whether $p_i$ should start a new round

\State $\mathtt{s}$, initially 0      \Comment Present round 

\State $\mathtt{suppout}(d,s)$, initially 0    \Comment Records whether responses from round $s$ support outputting $d$

\algstore{endvar}

\end{algorithmic}
\end{algorithm}

\vspace{0.2cm} 
\noindent \textbf{Pseudocode walk-through}:

\vspace{0.1cm} 
\noindent \textbf{Lines \ref{nrbegin}--\ref{nrend}}. The value $\mathtt{newround}$ indicates whether $p_i$ is ready to start a new round (and is initially set to 1). Lines \ref{nrbegin}--\ref{nrend} are carried out when $\mathtt{newround}=1$. They ask $p_i$ to sample a new set of processes, request their present colors, and record the start time for the present round, before setting $\mathtt{newround}=0$. 

\vspace{0.1cm} 
\noindent \textbf{Lines \ref{2Deltan}--\ref{updateend}}. These lines are responsible for updating the responses received from other processes. Since $p_i$ may proceed to  round $s+1$ before receiving sufficiently many responses from round $s$ to determine whether it should yet output, $p_i$ continues to collect responses for round $s$ after it has proceeded to later rounds. We write $x\uparrow$ to indicate that the variable $x$ is undefined, while $x\downarrow $ indicates that $x$ is defined. 

\vspace{0.1cm} 
\noindent \textbf{Lines \ref{updatesupportn}--\ref{updatesupportendn}}. 
The value $\mathtt{suppout}(d,s)$ indicates whether round $s$ supports finalizing $d$ ($\beta$ consecutive rounds of this type are required before outputting $d$). Lines \ref{updatesupportn}--\ref{updatesupportendn} are responsible for updating the values  $\mathtt{suppout}(d,s')$ for all $s'$ up to $p_i$'s present round. 

\vspace{0.1cm} 
\noindent \textbf{Lines \ref{lockupdatestartn}--\ref{lockupdateendn}}. The value $\mathtt{lock}$ (initially 0) indicates whether $p_i$ is locked on its present value. Process $p_i$ becomes locked upon seeing a new round in which it samples at least $\alpha_2$ of its present value. Process $p_i$ then becomes unlocked upon seeing a subsequent round in which it samples at least $\alpha_2$ processes that have been locked on the opposite color for at least $2\Delta$ (line \ref{31}). The value $\mathtt{lockbound}$ is used to ensure that only new samples cause $p_i$ to become freshly locked or unlocked. 

\vspace{0.1cm} 
\noindent \textbf{Lines \ref{readybegindn}--\ref{nr1dn}}. These lines test whether $p_i$ is ready to proceed to the next round. This is the case if it has received responses for the present round which already suffice to determine its color for the round, or if time $2\Delta$ has passed since starting the round. 

\vspace{0.1cm} 
\noindent \textbf{Lines \ref{f1d}--\ref{f2d}}. These lines determine whether $p_i$ is ready to output. Process $p_i$ outputs $d$ upon seeing $\beta$ consecutive rounds that support finalizing $d$. 

\vspace{0.1cm} 
\noindent \textbf{Lines \ref{resp1d}--\ref{resp2d}}. These lines have $p_i$ report its present color when requested, as well as how long it has been locked on that value.   

\setcounter{algorithm}{1}

\begin{algorithm} \label{pc:Snowflake}
\caption{Snowflake$^{\diamond}$: The instructions for  $p_i$ }
\begin{algorithmic}[1]

\algrestore{endvar}

  \State \textbf{At time $t$:}

  \State \hspace{0.3cm} \textbf{If} $\mathtt{newround}==1$:  \label{nrbegin} \Comment Start a new round if ready 

  \State \hspace{0.6cm} Form sample sequence $\langle p_{1,\mathtt{s}},\dots p_{k,\mathtt{s}} \rangle$;  \label{sample}         \Comment Sample  with replacement

  \State \hspace{0.6cm} For $j\in [1,k]$, send $\mathtt{s}$ to $p_{j,\mathtt{s}}$;    \Comment Ask  $p_{j,\mathtt{s}}$ for present value

  \State \hspace{0.6cm} Set $\mathtt{start}(\mathtt{s})=t$, $\mathtt{newround}:=0$; \Comment Set $t$ as start time for round $\mathtt{s}$ \label{nrend}

\State   

\State \hspace{0.3cm} For all $s'\in [0,\mathtt{s}]$ with $t>\mathtt{start}(s')\geq t-2\Delta$:  \label{2Deltan} \Comment Update values received 

\State \hspace{0.6cm} For all $j\in [1,k]$: 

\State \hspace{0.9cm} \textbf{If} $v_i(j,s')\uparrow$ and $p_i$ has received a message $(s',v,t_q)$ from $p_{j,s'}$: 

\State \hspace{1.2cm} Set $v_i(j,s'):=v$, $t_i(j,s')=t-t_q$; \label{updateend} \Comment Record value and locktime for $p_{j,s'}$ and round $s'$ 

\State \hspace{0.6cm} For  $d\in \{ 0,1 \}$: \label{updatesupportn}

\State \hspace{0.9cm} \textbf{If}  $|\{ j: 1\leq j \leq k, v_i(j,s')\downarrow==d, \ t_i(j,s')\leq \mathtt{start}(s')-2\Delta \}|\geq \alpha_2$:  \label{sfin}

\State \hspace{1.2cm} Set $\mathtt{suppout}(d,s'):=1$; \label{updatesupportendn} \Comment Record that values for round $s'$ support outputting $d$

\State

\State \hspace{0.3cm} \textbf{If} $\mathtt{lock}==0$:  \label{lockupdatestartn}  \Comment Update lock 

\State \hspace{0.6cm} \textbf{If} there exists a least $s'\geq \mathtt{lockbound}$ with

\State \hspace{0.6cm} $|\{ j: 1\leq j \leq k, v_i(j,s')\downarrow=\mathtt{val}_i \}|\geq \alpha_2 $ \textbf{and} such that 
 
 \State \hspace{0.6cm} for all $s'' \in [s',\mathtt{s})$, $\mathtt{val}_i(s'')==\mathtt{val}_i$:

\State \hspace{0.9cm} Set $\mathtt{lock}:=1$, $\mathtt{lockbound}:=s'+1$, $\mathtt{locktime}:=t$;




\label{lockupdateendn}

\State 

\State \hspace{0.3cm} \textbf{If} $\mathtt{lock}==0$: \label{readybegindn}  \Comment Look to see whether we should start a new round

\State \hspace{0.6cm} \textbf{If} $|\{ j: 1\leq j \leq k, v_i(j,\mathtt{s})==\mathtt{val}_i \}| \geq k-\alpha_1+1$: 

\State \hspace{0.9cm} Set $\mathtt{val}_i(\mathtt{s}):=\mathtt{val}_i$, $\mathtt{s}:=\mathtt{s}+1$, $\mathtt{newround}:=1$;  \label{nr1} 

\State \hspace{0.6cm} \textbf{If} $|\{ j: 1\leq j \leq k, v_i(j,\mathtt{s})==1-\mathtt{val}_i \}| \geq \alpha_1$: 

\State \hspace{0.9cm} Set $\mathtt{val}_i:=1-\mathtt{val}_i$, $\mathtt{val}_i(\mathtt{s}):=\mathtt{val}_i$, $\mathtt{s}:=\mathtt{s}+1$, $\mathtt{newround}:=1$; 

\State \hspace{0.3cm} \textbf{If} $\mathtt{lock}==1$: 

\State \hspace{0.6cm} \textbf{If} $|\{ j: 1\leq j \leq k, v_i(j,\mathtt{s})==\mathtt{val}_i \lor t_i(j,\mathtt{s})>\mathtt{start}(s)-2\Delta \}| \geq k-\alpha_2+1$: 

\State \hspace{0.9cm} Set $\mathtt{val}_i(\mathtt{s}):=\mathtt{val}_i$, $\mathtt{s}:=\mathtt{s}+1$, $\mathtt{newround}:=1$;  \label{nr1} 

\State \hspace{0.6cm} \textbf{If} $|\{ j: 1\leq j \leq k, v_i(j,\mathtt{s})==1-\mathtt{val}_i \land t_i(j,\mathtt{s}) \leq \mathtt{start}(s)-2\Delta  \}| \geq \alpha_2$: \label{sunlock}

\State \hspace{0.9cm} Set $\mathtt{val}_i:=1-\mathtt{val}_i$, $\mathtt{val}_i(\mathtt{s}):=\mathtt{val}_i$, $\mathtt{s}:=\mathtt{s}+1$, $\mathtt{newround}:=1$;
\State \hspace{0.9cm} Set $\mathtt{lock}:=0$ and make $\mathtt{locktime}$ undefined; \label{31} \Comment Unlock

\State \hspace{0.3cm} \textbf{If} $\mathtt{start}(\mathtt{s})\downarrow \leq t-2\Delta$: \Comment Proceed to next round on timeout

\State \hspace{0.6cm} Set $\mathtt{val}_i(\mathtt{s}):=\mathtt{val}_i$, $\mathtt{s}:=\mathtt{s}+1$, $\mathtt{newround}:=1$;  \label{nr1dn}

\State 

\State \hspace{0.3cm} For $d\in \{ 0,1 \}$: \label{f1d} \Comment See whether to finalize 

\State \hspace{0.6cm} \textbf{If} there exists $s'$ such that, for all $s''\in [s',s'+\beta)$, $\mathtt{suppout}(d,s'')==1$:  \label{fincon0} 
\State \hspace{0.9cm} Output $d$ and terminate;  \label{f2d}

\State

\State \hspace{0.3cm} \textbf{If} $p_i$ has received $s$ from $p_j$ (for the first time):  \label{resp1d} \Comment Send requested values

\State \hspace{0.6cm} \textbf{If} $\mathtt{lock}==0$:

\State \hspace{0.9cm} Send $(s,\mathtt{val}_i,0)$ to $p_j$;

\State \hspace{0.6cm} \textbf{If} $\mathtt{lock}==1$:

\State \hspace{0.9cm} Send $(s,\mathtt{val}_i,t-\mathtt{locktime})$ to $p_j$; \label{resp2d}

\end{algorithmic}
\end{algorithm}

\section{Security analysis of Snowflake$^{\diamond}$}       \label{SnowFDanal}

 We consider the partially synchronous setting, assume $f<n/5$ and establish satisfaction of agreement for $k=80$, $\alpha_1=41$, $\alpha_2=72$, and $\beta=12$, under the assumption that the population size $n\geq 250$. We note that the bound $f$ is on the number of Byzantine processes: since we only consider agreement in partial synchrony, our analysis will tolerate an arbitrary number of crash or omission failures. The structure of the argument is similar to that presented in Section \ref{Snowflakeanalysis}. We make the assumption that $f<n/5$ and $n\geq 250$ only so as to be able to give as simple a proof as possible: a more fine-grained analysis for smaller $n$ is the subject of future work. 

\vspace{0.2cm} Note that the last action a process performs at each timeslot is to report its present value to others (lines \ref{resp1d}--\ref{resp2d}). We say $p_i$ is blue/red at $t$ if $\mathtt{val}_i$ is 0/1 at the end of timeslot $t$ (which must then be the color it reports to others at $t$ if correct). Similarly, we say 
$p_i$ is \emph{locked on blue/red} at $t$ if $\mathtt{val}_i$ is 0/1 and its local value $\mathtt{lock}$ equals 1 at the end of timeslot $t$. We say $p_i$ \emph{outputs blue/red} at $t$ if it outputs $0/1$ at $t$.

\vspace{0.2cm}
\noindent \textbf{Establishing Agreement}. As in Section \ref{Snowflakeanalysis}, the arguments consists of three parts.

\vspace{0.2cm} 
\noindent \textbf{Part 1}. We begin by considering  what happens when the proportion of correct processes that are locked on red reaches a certain threshold. In particular, let us consider what happens when $t$ satisfies the property that at least 75\% of the correct  processes are locked on red at $t$. If $p_i$ starts a new round $s$ and samples the sequence  $\langle p_{1,s},\dots p_{k,s} \rangle$ at  $t$ (line \ref{sample}), then a calculation for the binomial shows that the probability that at most 8 (i.e.\ $k-\alpha_2$) members of this sequence are correct and locked on red at $t$ is $<1.18\times 10^{-20}$, i.e.\ $\text{Bin}(80,0.8\times 0.75,\leq 8)<1.18 \times 10^{-20}$.

The analysis above applies to any single given round $s$ and a single process. Next, we wish to bound the probability that the following statement holds for all $t$ and all processes: 
\begin{enumerate} 
\item[$(\dagger_0^{\diamond})$] Suppose at least 75\% of the correct processes are locked on red at $t$.  If a correct process starts a new round $s$ and samples the sequence  $\langle p_{1,s},\dots p_{k,s} \rangle$ at  $t$, then at least 9 members of this sequence are correct and locked on red at $t$. 
\end{enumerate}
\noindent  To understand the significance of $(\dagger_0^{\diamond})$, note that if $p_i$ starts a new round $s$ and samples the sequence  $\langle p_{1,s},\dots p_{k,s} \rangle$ at  $t$, and if $p_{j,s}$ is correct and locked on red at $t$, then $p_{j,s}$ cannot respond to $p_i$'s request to send values for round $s$ by reporting (before $p_i$ stops collecting values for that round after time $2\Delta$) that it has been locked on blue for at least $2\Delta$. To bound the probability that $(\dagger_0^{\diamond})$ fails to hold, we can bound the number of processes and the number of rounds initiated by any process, and then apply the union bound to our analysis above. Suppose that the protocol is executed for at most 1000 years by at most 10000 processes, with each process initiating at most 5 rounds per second.  This means that each process initiates less than $1.6\times 10^{11}$ rounds. The union bound thus gives a cumulative error probability of less than $(1.6\times 10^{11}) \times 10^4 \times (1.18\times 10^{-20})  < 1.9\times 10^{-5}$, meaning that $(\dagger_0^{\diamond})$ fails to hold  with probability at most  $1.9\times 10^{-5}$.

\vspace{0.2cm} 
Extending this idea, suppose next that at least 75\% of the correct processes are locked on red for the entirety of the interval $[t,t+2\Delta]$. Let $P$ be the set of correct processes that are locked on red for the entirety of this interval. We aim to show that all processes in $P$ remain locked on red thereafter (except with small probability). Towards a contradiction, suppose there is some first timeslot $t^*>t+2\Delta$ at which $p_i\in P$ ceases to be locked on red, after receiving sufficiently many values for some round $s$ (line \ref{31}). Since $p_i$ ceases to update values $v_i(j,s)$ time $2\Delta$ after starting round $s$  (see line \ref{2Deltan}), $p_i$ must start round $s$ at some timeslot $t_0$ in the interval $[t^*-2\Delta, t^*)$. Since all members of $P$ are locked on red at $t_0$ (by our choice of $t^*$), $(\dagger_0^{\diamond})$ together with the observation that follows its statement above give an immediate contradiction. It therefore follows from $(\dagger_0^{\diamond})$ that: 

\begin{enumerate} 
\item[$(\dagger_1^{\diamond})$] Suppose that $P$ is a set of correct processes and that all members of $P$ are locked on red for the entirety of the interval $[t,t+2\Delta]$. If at least 75\% of correct processes are in $P$, then every member of $P$ remains locked on red at all timeslots $\geq t$. 
\end{enumerate}



\vspace{0.2cm} 
\noindent \textbf{Part 2}.
Next, we establish our analogue of $(\dagger_2)$. Let $\mathtt{start}(s)$ be the timeslot at which $p_i$ begins round $s$, and let the sequence  $\langle p_{1,s},\dots p_{k,s} \rangle$  be as locally defined for $p_i$. If at most 75\% correct processes are locked on red for the entirety of the interval $[\mathtt{start}(s)-2\Delta,\mathtt{start}(s)]$, then a calculation for the binomial distribution shows that the probability at least  72 or more members of the sequence  $\langle p_{1,s},\dots p_{k,s} \rangle$ are Byzantine or locked on red for the entirety of the interval $[\mathtt{start}(s)-2\Delta,\mathtt{start}(s)]$ is upper bounded by 0.0131, i.e.\ $\text{Bin}(80,(0.75\times 0.8)+0.2,\geq 72)<0.0131$. If, for some $x\geq 1$ it then holds that at most 75\% correct processes are locked on red for the entirety of the interval $[\mathtt{start}(s+x)-2\Delta,\mathtt{start}(s+x)]$, then (independent of previous events), the probability  at least  72 or more members of the sequence  $\langle p_{1,s+x},\dots p_{k,s+x} \rangle$ are Byzantine or locked on red for the entirety of the interval $[\mathtt{start}(s+x)-2\Delta,\mathtt{start}(s+x)]$ is again upper bounded by 0.0131. 
So, now consider any 12 given consecutive rounds, those in the interval $[s,s+\beta)$ say,  and any given correct process $p_i$. For $s'\in [s,s+\beta)$, let $\mathtt{suppout}(1,s')$  be as locally defined for $p_i$.  The probability that $\mathtt{suppout}(1,s')=1$ for all $s'\in [s,s+\beta)$ and there does not exist any interval $[t'-2\Delta,t']$ with $t'\leq \mathtt{start}(s+\beta-1)$ such that at least 75\% of correct processes are locked on red for the entirety of the interval $[t'-2\Delta,t']$ is upper bounded by  $0.0131^{12}<10^{-22}$. If at most 10,000 processes execute the protocol for at most 1000 years, executing at most 5 rounds per second, we can then apply the union bound to conclude that the following statement fails to hold (throughout the execution) with probability at most $10^{-22}\times 10000 \times (1.6 \times 10^{11})<  2\times 10^{-7}$:

\begin{enumerate}
\item[$(\dagger_2^{\diamond})$] If a correct process outputs red at $t$, then, for some $t'\leq t$, at least 75\% of correct processes are locked on red for the entirety of the interval $[t'-2\Delta,t']$. 
\end{enumerate}

\vspace{0.2cm} 
\noindent \textbf{Part 3}. Now we put parts 1 and 2 together. From the union bound and the analysis above, we may conclude that $(\dagger_1^{\diamond})$ and $(\dagger_2^{\diamond})$ both hold, except with probability at most 
$(1.9\times 10^{-5})+( 2\times 10^{-7})<2\times 10^{-5}$. So, suppose these conditions both hold. According to $(\dagger_2^{\diamond})$,  if a correct process is the (potentially joint) first to output and outputs  red at $t$, then there exists $t'\leq t$ such that at least 75\% of correct processes are locked on red for the entirety of the interval $[t'-2\Delta,t']$. From $(\dagger_1^{\diamond})$ it follows that, for all $t''\geq t'$,   at least 75\% of correct processes are locked on red for the entirety of the interval $[t'-2\Delta,t'']$.  From $(\dagger_2^{\diamond})$ (applied to blue), it follows that no correct process ever outputs blue. This suffices to show that agreement is satisfied, except with small  error probability.

\section{Snowman$^{\diamond}$} \label{SnowmanD}

 We build Snowman$^{\diamond}$ from Snowflake$^{\diamond}$ using a similar approach to that used in \cite{buchwald2024frosty}  to build Snowman from Snowflake$^+$. As for Snowflake$^{\diamond}$, Snowman$^{\diamond}$ operates in the partially synchronous setting and does not assume that clocks are synchronized. 

\subsection{Transactions and blocks}
To specify a protocol for State-Machine-Replication (SMR), we suppose processes are sent (signed) transactions during the protocol execution: Formally this can be modeled by having processes be sent transactions by an \emph{environment}, e.g.\ as in \cite{lewis2023permissionless}. Processes may use received transactions to form \emph{blocks} of transactions. To make the analysis as general as possible, we decouple the process of block production from the core consensus engine. We therefore suppose that some given process for block generation operates in the background, and that valid blocks are gossiped throughout the network. We do not put constraints on the block generation process, and allow that it may produce equivocating blocks, etc. In practice, block generation could be specified simply by having a rotating sequence of leaders propose blocks, or through a protocol such as Snowman$^{++}$, as actually used by the present implementation of the Avalanche blockchain (for a description of Snowman$^{++}$, see \cite{plusplus}). 

\vspace{0.2cm} 
\noindent \textbf{Blockchain structure}. We consider a fixed \emph{genesis block} $b_0$. In a departure from the approach described in the original Avalanche whitepaper \cite{rocket2019scalable}, which built a directed acyclic graph (DAG) of blocks, we consider a standard linear blockchain architecture in which each block $b$ other than $b_0$ specifies a unique \emph{parent}.  
If $b'$ is the parent of $b$, then $b$ is referred to as a \emph{child} of $b'$. In this case, the ancestors of $b$ are $b$
and any ancestors of $b'$. Every block must have $b_0$ as an ancestor. The descendants of any block $b$
are $b$ and any descendants of its children.  The \emph{height} of a block $b$ is its number of ancestors other than $b$, meaning that the height of $b_0$ is 0. 
By a \emph{chain} (ending in $b_h$), we mean a sequence of blocks $b_0 \ast b_1 \ast \dots \ast b_h$, such that $b_{h'+1}$ is a child of $b_{h'}$ for $h'<h$.\footnote{Throughout this paper, `$\ast$' denotes concatenation.} 


\subsection{Overview of the Snowman$^{\diamond}$ protocol}

To implement SMR, our approach is to run multiple instances of Snowflake$^{\diamond}$. 
To keep things simple, consider first the task of reaching consensus on a block of height 1. Suppose that multiple children of $b_0$ are proposed over the course of the execution and that we must choose between them. To turn this decision problem into multiple binary decision problems,  we consider the hash value $H(b_1)$ of each proposed block $b_1$ of height 1, and then run one instance of Snowflake$^{\diamond}$ to reach consensus on the first bit of the hash. Then we run  a second instance to reach consensus on the second bit of the hash, and so on. Working above a block of any height $h$, the same process is then used to finalize a block of height $h+1$. In this way, multiple instances of Snowflake$^{\diamond}$ are used to reach consensus on a chain of hash values $H(b_0) \ast H(b_1) \ast \dots$. 

\vspace{0.2cm} 
This process would not be efficient if each round required a separate set of  correspondences for each instance of Snowflake$^{\diamond}$, but this is not necessary. Just as in Snowflake$^{\diamond}$, at the beginning of each round $s$, process $p_i$ samples a single sequence $\langle p_{1,s},\dots p_{k,s} \rangle$ of $k$ processes. Since we now wish to reach consensus on a sequence of blocks, each process $p_{j,s}$ in the sample is now requested to report its presently preferred chain, rather than a single bit value. The first bit of the corresponding hash sequence is then used by $p_i$ as the response of $p_{j,s}$ in a first instance of Snowflake$^{\diamond}$. If this first bit agrees with $p_i$'s resulting value in that instance of Snowflake$^{\diamond}$, then the second bit  is used as the value reported by $p_{j,s}$ in a second instance of Snowflake$^+$, and so on. 
Proceeding in this way, Snowflake is used to decide between competing hash values. To decide between two hashes, we only need a single instance of Snowflake$^{\diamond}$, which is used to decide the first bit that the two hashes disagree on. 


\vspace{0.2cm} 
\noindent \textbf{A note on some simplifications that are made for the sake of clarity of presentation}.  When a process $p_{j,s}$ is requested by $p_i$ to report its presently preferred chain (ending with $b$, say), we have $p_{j,s}$ simply send the given sequence of blocks. In reality, this would be very inefficient and the present implementation of Snowman deals with this by having $p_{j,s}$ send a hash of $b$ instead. This potentially causes some complexities, because $p_i$ may not have seen  $b$ (meaning that it does not necessarily know how to interpret the hash). This issue is easily dealt with, but it would be a distraction to go into the details here.

\vspace{0.2cm} 
\noindent \textbf{The variables, functions and procedures used by $p_i$}. The protocol instructions make use of the following variables and functions (as well as others whose use should be clear from the pseudocode):

\begin{itemize}
\item $b_0$: The genesis block.
\item $\mathtt{blocks}$: Stores blocks received by $p_i$ (and verified as valid). Initially it contains only $b_0$, and it is automatically updated over time to include any block included in any message received or sent by $p_i$.
\item $\mathtt{val}(\sigma)$, initially undefined: For each finite binary string $\sigma$, $\mathtt{val}(\sigma)$ records $p_i$'s presently preferred value for the next bit of the chain of hash values $H(b_0) \ast H(b_1) \ast \dots$, should the latter extend $\sigma$. 
\item $\mathtt{pref}$, initially set to $H(b_0)$: The initial segment of the chain of hash values that $p_i$ presently prefers. We write $|\mathtt{pref}|$ to denote the length of this binary string.
\item $\mathtt{pref}(s)$, initially undefined: $p_i$'s $\mathtt{pref}$ value at the end of round $s$. 
\item $\mathtt{final}$, initially set to $H(b_0)$: The initial segment of the chain of hash values that $p_i$ presently regards as final. 

\item $\mathtt{chain}(\sigma)$: If there exists a greatest $h\in \mathbb{N}$ such that $\sigma=H(b_0) \ast \dots \ast H(b_h) \ast \tau$ for a chain of blocks $b_0 \ast \dots \ast b_h$ all seen by $p_i$, and for some finite string $\tau$, then $\mathtt{chain}(\sigma):= b_0 \ast \dots \ast b_h$. Otherwise, $\mathtt{chain}(\sigma):=b_0$. 

\item $\mathtt{reduct}(\sigma)$: If there exists a greatest $h\in \mathbb{N}$ such that $\sigma=H(b_0) \ast \dots \ast H(b_h) \ast \tau$ for a chain of blocks $b_0 \ast \dots \ast b_h$ all seen by $p_i$, and for some finite string $\tau$, then $\mathtt{reduct}(\sigma):=H(b_0) \ast \dots \ast H(b_h)$. Otherwise, $\mathtt{reduct}(\sigma):=H(b_0)$. 

\item $\mathtt{last}(\sigma)$: If there exists a greatest $h\in \mathbb{N}$ such that $\sigma=H(b_0) \ast \dots \ast H(b_h) \ast \tau$ for a chain  $b_0 \ast \dots \ast b_h$ all seen by $p_i$, and for some finite string $\tau$, then $\mathtt{last}(\sigma):= b_h$. Otherwise, $\mathtt{last}(\sigma):=b_0$.
\item $H_B$: If $B=b_0\ast b_1 \ast \dots \ast b_h$ is a chain, then $H_B:=H(b_0) \ast H(b_1) \ast \dots H(b_h)$, and if not then $H_B$ is the empty string $\emptyset$.
\item $\mathtt{start}(s)$, initially undefined: The time at which $p_i$ starts round $s$ (according to local clock).

\item  $\mathtt{lock}(\sigma)$, initially 0: Indicates whether $p_i$ is locked on $\sigma$.

\item $\mathtt{locktime}(\sigma)$, initially undefined: The time at which $p_i$ became locked on $\sigma$.

\item $\mathtt{lockbound}(\sigma)$, initially 0: Rounds after this may cause $p_i$ to become freshly locked.

\item $\mathtt{newround}$, initially 1: Indicates whether $p_i$ should start a new round.

\item  $\mathtt{s}$, initially 0: Present round. 

\item $\mathtt{suppfin}(\sigma,s)$, initially 0: Records whether responses from round $s$ support finalizing $\sigma$.

\item $\mathtt{dec}(s,\sigma)$, initially 0: Records whether values for round $s$ already suffice to determine that $\sigma$ is an initial segment of $\mathtt{pref}(s)$.

\item $\mathtt{rpref}(j,s)$, initially undefined: The preferred chain that $p_{j,s}$ reports to $p_i$ in round $s$. 

\item $\mathtt{rlock}(j,s)$, initially undefined: The initial segment of $\mathtt{rpref}(j,s)$ that $p_{j,s}$ reports to $p_i$ as having been locked for time $4\Delta$. 
   
\end{itemize}

The pseudocode is shown in Algorithm \ref{pc:Snowmand}.  For strings $\sigma$ and $\tau$, we write $\sigma \subseteq \tau$ to denote that (both of these values are defined and) $\sigma$ is an initial segment of $\tau$.

\vspace{0.2cm} 
\noindent \textbf{Pseudocode walk-through}:

\vspace{0.1cm} 
\noindent \textbf{Lines \ref{nrbegind}--\ref{nrendd}}. The value $\mathtt{newround}$ indicates whether $p_i$ is ready to start a new round (and is initially set to 1). Lines \ref{nrbegind}--\ref{nrendd} are carried out when $\mathtt{newround}=1$. They ask $p_i$ to sample a new set of processes, request their present colors, and record the start time for the present round, before setting $\mathtt{newround}=0$. 

\vspace{0.1cm} 
\noindent \textbf{Lines \ref{2Deltad}--\ref{updateendd}}. These lines are responsible for updating the responses received from other processes. Since $p_i$ may proceed to  round $s+1$ before receiving sufficiently many responses from round $s$ to determine whether it should yet output, $p_i$ continues to collect responses for round $s$ after it has proceeded to later rounds. We write $x\uparrow$ to indicate that the variable $x$ is undefined, while $x\downarrow $ indicates that $x$ is defined. 

\vspace{0.1cm} 
\noindent \textbf{Lines \ref{updatesupportd}--\ref{updatesupportendd}}. 
The value $\mathtt{suppfin}(\sigma,s)$ indicates whether round $s$ supports finalizing $\sigma$ ($\beta$ consecutive rounds of this type are required before finalizing $\sigma$). Lines \ref{updatesupportd}--\ref{updatesupportendd} are responsible for updating the values  $\mathtt{suppfin}(\sigma,s')$ for all $s'$ up to $p_i$'s present round and which started at most $2\Delta$ ago. 

\vspace{0.1cm} 
\noindent \textbf{Lines \ref{lockupdatestartd}--\ref{lockupdateendd}}. The value $\mathtt{lock}(\sigma)$ (initially 0) indicates whether $p_i$ is locked on $\sigma$. Process $p_i$ becomes locked on $\sigma$ upon seeing a new round in which it samples at least $\alpha_2$ values extending $\sigma$. Process $p_i$ then becomes unlocked upon seeing a subsequent round in which it samples at least $\alpha_2$ processes that report having been locked on a value incompatible with $\sigma$ for time at least $4\Delta$, meaning time at least $2\Delta$ from the point at which $p_i$ starts that round (line \ref{41}). The value $\mathtt{lockbound}$ is used to ensure that only new samples cause $p_i$ to become freshly locked. 

\vspace{0.1cm} 
\noindent \textbf{Lines \ref{21}--\ref{45}}. These lines iterate up from $p_i$'s present value $\mathtt{final}$ to recalculate $\mathtt{pref}$, based on responses from the present round, and also test whether $p_i$ is ready to proceed to the next round. This is the case if it has received responses for the present round which already suffice to determine $\mathtt{pref}(s)$, or if time $2\Delta$ has passed since starting the round. The value $\mathtt{dec}(\mathtt{s},\sigma)$ (intially 0) specifies whether responses for the present round $\mathtt{s}$ already suffice to determine that $\sigma$ will be an initial segment of $\mathtt{pref}(\mathtt{s})$ ($p_i$'s $\mathtt{pref}$ value at the end of round $\mathtt{s}$). 

\vspace{0.1cm} 
\noindent \textbf{Lines \ref{final?}--\ref{f3}}. These lines determine whether $p_i$ is ready to extend $\mathtt{final}$. 

\vspace{0.1cm} 
\noindent \textbf{Lines \ref{resps}--\ref{resps2}}. These lines have $p_i$ report its present values when requested.

\begin{algorithm} 
\caption{Snowman$^{\diamond}$: The instructions for  $p_i$ }  \label{pc:Snowmand}
\begin{algorithmic}[1]
\scriptsize

  \State \textbf{At time $t$:}

  \State \hspace{0.3cm} \textbf{If} $\mathtt{newround}==1$:  \label{nrbegind} \Comment Start a new round if ready 

  \State \hspace{0.6cm} Form sample sequence $\langle p_{1,\mathtt{s}},\dots p_{k,\mathtt{s}} \rangle$;  \label{sampled}         \Comment Sample  with replacement

  \State \hspace{0.6cm} For $j\in [1,k]$, send $\mathtt{s}$ to $p_{j,\mathtt{s}}$;  \label{ask}  \Comment Ask  $p_{j,\mathtt{s}}$ for present value

  \State \hspace{0.6cm} Set $\mathtt{start}(\mathtt{s})=t$, $\mathtt{newround}:=0$; \Comment Set $t$ as start time for round $\mathtt{s}$ \label{nrendd}

\State   

\State \hspace{0.3cm} For all $s'\in [0,\mathtt{s}]$ with $t>\mathtt{start}(s')\geq t-2\Delta$:  \label{2Deltad} \Comment Update values received 

\State \hspace{0.6cm} For all $j\in [1,k]$: 

\State \hspace{0.9cm} \textbf{If} $\mathtt{rpref}(j,s')\uparrow$ and $p_i$ has received a message $(s',B,\sigma)$ from $p_{j,s'}$ s.t.\ $B$ is a chain
  with $\sigma \subseteq H_B$: 

\State \hspace{1.2cm} Set $\mathtt{rpref}(j,s'):=H_B$, $\mathtt{rlock}(j,s')=\sigma$; \label{updateendd} \Comment Record values for $p_{j,s'}$ and round $s'$ 

\State \hspace{0.6cm} For  all $\sigma\subseteq \mathtt{pref}$: \label{updatesupportd}

\State \hspace{0.9cm} \textbf{If}  $|\{ j: 1\leq j \leq k, \mathtt{rlock}(j,s')\downarrow\supseteq \sigma \}|\geq \alpha_2$:  \label{sfin}

\State \hspace{1.2cm} Set $\mathtt{suppfin}(\sigma,s'):=1$; \label{updatesupportendd} \Comment Record that values for round $s'$ support finalizing $\sigma$

\State

\State \hspace{0.3cm} For all $\sigma \subseteq \mathtt{pref}$, \textbf{if} $\mathtt{lock}(\sigma)==0$:  \label{lockupdatestartd}  \Comment Update locks 

\State \hspace{0.6cm} \textbf{If} there exists a least $s'\geq \mathtt{lockbound}(\sigma)$ 
such that

\State \hspace{0.6cm}  $|\{ j: 1\leq j \leq k, \mathtt{rpref}(j,s')\supseteq \sigma  \}|\geq \alpha_2 $ \textbf{and} such that

\State \hspace{0.6cm} for all $s'' \in [s',\mathtt{s})$, $\mathtt{pref}(s'')\supseteq \sigma$:

\State \hspace{0.9cm} Set $\mathtt{lock}(\sigma):=1$, $\mathtt{lockbound}(\sigma):=s'+1$, $\mathtt{locktime}(\sigma):=t$;




\label{lockupdateendd}

  \State 

   \State \hspace{0.3cm} Set $\mathtt{pref}:=\mathtt{final}$, $\mathtt{end}:=0$; \label{21} \Comment{Begin iteration to update $\mathtt{pref}$} 

  \State \hspace{0.3cm} \textbf{While} $\mathtt{end}==0$ \textbf{do}: 

  \State \hspace{0.6cm} Set $E:= \{ b\in \mathtt{blocks}:\ b \text{ is a child of }\mathtt{last}(\mathtt{pref}) \text{ and }\mathtt{pref} \subseteq \mathtt{reduct}(\mathtt{pref})\ast H(b)\}$;

  \State \hspace{0.6cm} \textbf{If} $E$ is empty, set $\mathtt{end}:=1$;

 \State \hspace{0.6cm} \textbf{Else}:        \Comment{Carry out the next instance of Snowflake$^\diamond$}

 \State \hspace{0.9cm} \textbf{If} $\mathtt{val}(\mathtt{pref})$ is undefined:

\State \hspace{1.2cm} Let $b$ be the first block in $E$ enumerated into $\mathtt{blocks}$;  

\State \hspace{1.2cm} Set $\mathtt{val}(\mathtt{pref})$ to be the $(|\mathtt{pref}|+1)^{\text{th}}$ bit of $\mathtt{reduct}(\mathtt{pref})\ast H(b)$;

\State \hspace{0.9cm} \textbf{If} $\mathtt{lock}(\mathtt{pref}\ast \mathtt{val}(\mathtt{pref}))==0$: \label{readybegin}  \Comment Look to see whether we can decide next bit of $\mathtt{pref}(\mathtt{s})$

\State \hspace{1.2cm} \textbf{If} $|\{ j: 1\leq j \leq k, \mathtt{rpref}(j,\mathtt{s})\downarrow \not \supseteq \mathtt{pref} \ast (1-\mathtt{val}(\mathtt{pref})) \}| \geq k-\alpha_1+1$: 

\State \hspace{1.5cm} Set $\mathtt{dec}(\mathtt{s},\mathtt{pref} \ast \mathtt{val}(\mathtt{pref} )):=1$;  \label{nr1} 

\State \hspace{1.2cm} \textbf{If} $|\{ j: 1\leq j \leq k, \mathtt{rpref}(j,\mathtt{s})\supseteq \mathtt{pref} 
\ast 1-\mathtt{val}(\mathtt{pref}) \}| \geq \alpha_1$: 

\State \hspace{1.5cm} Set $\mathtt{val}(\mathtt{pref}):=1-\mathtt{val}(\mathtt{pref})$;
\State \hspace{1.5cm}  Set $\mathtt{dec}(\mathtt{s},\mathtt{pref} \ast \mathtt{val}(\mathtt{pref} )):=1$;

\State \hspace{0.9cm} \textbf{If} $\mathtt{lock}(\mathtt{pref}\ast \mathtt{val}(\mathtt{pref}))==1$: 

\State \hspace{1.2cm} \textbf{If} $|\{ j: 1\leq j \leq k, \mathtt{rlock}(j,\mathtt{s})\downarrow \not \supseteq \mathtt{pref} \ast (1-\mathtt{val}(\mathtt{pref})) \}| \geq k-\alpha_2+1$: 

\State \hspace{1.5cm} Set $\mathtt{dec}(\mathtt{s},\mathtt{pref} \ast \mathtt{val}(\mathtt{pref} )):=1$;

\State \hspace{1.2cm} \textbf{If} $|\{ j: 1\leq j \leq k, \mathtt{rlock}(j,\mathtt{s})\supseteq \mathtt{pref} 
\ast 1-\mathtt{val}(\mathtt{pref}) \}| \geq \alpha_2$:  \label{sunlock}

\State \hspace{1.5cm} Set $\mathtt{val}(\mathtt{pref}):=1-\mathtt{val}(\mathtt{pref})$;
\State \hspace{1.5cm}  Set $\mathtt{dec}(\mathtt{s},\mathtt{pref} \ast \mathtt{val}(\mathtt{pref} )):=1$;

\State \hspace{1.5cm} For all $\sigma \supset \mathtt{pref}$, set $\mathtt{lock}(\sigma):=0$ and make $\mathtt{locktime}(\sigma)$ undefined; \label{41} \Comment Unlock

 \State \hspace{0.6cm} Set $\mathtt{pref}:=\mathtt{pref} \ast \mathtt{val}(\mathtt{pref})$; \label{endpref}

\State

\State \hspace{0.3cm} \textbf{If} $\mathtt{start}(\mathtt{s})\downarrow \leq t-2\Delta$ \textbf{or} $\mathtt{dec}(\mathtt{s},\sigma)==1$ for all  $\sigma$ with $\mathtt{final}\subset \sigma \subseteq \mathtt{pref}$: 

\State \hspace{0.6cm} Set $\mathtt{pref}(\mathtt{s}):=\mathtt{pref}$, $\mathtt{s}:=\mathtt{s}+1$, $\mathtt{newround}:=1$;  \label{45} \Comment Proceed to next round

\label{nr2}

\State 

\State \hspace{0.3cm} \textbf{If} there exists a longest $\sigma$ s.t.\ $\mathtt{final} \subset \sigma \subseteq \mathtt{pref}$ \textbf{and} such that there exists $s'$ such that, \label{final?}

\State \hspace{0.3cm}  for all $s''\in [s',s'+\beta)$, $\mathtt{suppfin}(\sigma,s'')==1$:  \label{fincon2} 
\State \hspace{0.6cm} Set  $\mathtt{final}:=\sigma$;  \label{f3}

\State

\State \hspace{0.3cm} \textbf{If} $p_i$ has received $s$ from $p_j$ (for the first time):  \label{resps} \Comment Send requested values

\State \hspace{0.6cm} Let $\sigma$ be the longest initial segment of $\mathtt{pref}$ s.t. $t-\mathtt{locktime}(\sigma)\downarrow \geq 4\Delta$; \label{3D}

\State \hspace{0.6cm} Send $(s,\mathtt{chain}(\mathtt{pref}),\sigma)$ to $p_j$; \label{resps2}

\end{algorithmic}
\end{algorithm}

\section{Consistency analysis for Snowman$^{\diamond}$} \label{consisanal}
We write $\mathtt{pref}_i$ and $\mathtt{final}_i$ to denote the values $\mathtt{pref}$ and $\mathtt{final}$ as locally defined for $p_i$. We say \emph{$p_i$ finalizes $\sigma$}, or $\sigma$ \emph{becomes final for $p_i$}, if there exists some round during which $\sigma \subseteq \mathtt{final}_i$.  

\vspace{0.2cm} 
\noindent \textbf{Consistency}: Suppose $\sigma:=\mathtt{final}_i$ as defined at the beginning of round $s$ and that 
$\sigma':=\mathtt{final}_j$  as defined at the beginning of round $s'$. Then, whenever $p_i$ and $p_j$ are correct: 
\begin{itemize} 
\item[(i)] If $i=j$ and $s'\geq s$ then $\sigma \subseteq \sigma'$. 
\item[(ii)] Either $\sigma$ extends $\sigma'$, or $\sigma'$ extends $\sigma$. 
\end{itemize}

In this section, we show that Snowman$^{\diamond}$ satisfies consistency in partial synchrony (except with small  error probability) for appropriate choices of the protocol parameters, and so long as $f<n/5$. Just as for Snowflake$^{\diamond}$, the analysis holds for any number of crash or omission failures. As in previous sections, for the sake of concreteness we give an analysis for $k=80$, $\alpha_1=41$, $\alpha_2=72$, and $\beta=12$, under the assumption that the population size $n\geq 250$. We make the assumption that $f<n/5$ and $n\geq 250$ only so as to be able to give as simple a proof as possible: a more fine-grained analysis for smaller $n$ is the subject of future work. 
.

\vspace{0.2cm} 
\noindent \textbf{The proof of consistency}. It follows directly from the protocol instructions that (i) in the definition of consistency is satisfied. To see this, note that, initially, $\mathtt{pref}_i=\mathtt{final}_i=H(b_0)$. The values $\mathtt{pref}_i$ and $\mathtt{final}_i$ are not redefined during round $s$ prior to line \ref{21}, when we set $\mathtt{pref}_i:=\mathtt{final}_i$. If $\mathtt{pref}_i$ is subsequently redefined during round $s$, then we redefine it to be an extension of its previous value. If $\mathtt{final}_i$ is redefined during round $s$, then it is defined to be an extension of the present value of $\mathtt{pref}_i$. 

\vspace{0.2cm} To argue that (ii) in the definition of consistency is satisfied,  we suppose again that the protocol is run by at most 10,000 processes for at most 1000 years, each process executing at most 5 rounds per second.  We'll say a correct process $p_i$ `samples $x$ values extending $\sigma$' in round $s$ if
$|\{ j: 1\leq j \leq k, \mathtt{rpref}(j,s)  \supseteq \sigma  \}|=x$,   where $\mathtt{rpref}(j,s)$ is as locally defined for $p_i$ at the end of round $s$.  We'll say a correct process $p_i$ `samples $x$ locked values extending $\sigma$' in round $s$ if
$|\{ j: 1\leq j \leq k, \mathtt{rlock}(j,s)  \supseteq \sigma  \}|=x$,   where $\mathtt{rlock}(j,s)$ is as locally defined for $p_i$ at the end of round $s$. We'll say $p_i$ is `locked on $\sigma$' at $t$ if its local value $\mathtt{lock}(\sigma)=1$ at the end of timeslot $t$. 
We define $\sigma_t$ to be the longest string $\sigma$ such that at least 75\% of correct processes have been locked on extensions of $\sigma$ for at least $2\Delta$ at $t$, and we define $\sigma_t^*$ to be the longest string $\sigma$ such that at least 75\% of correct processes are locked on extensions of $\sigma$ at $t$.

\vspace{0.2cm} 
The argument consists of three parts, similar to those described in Section \ref{Snowflakeanalysis}.

\vspace{0.2cm} 
\noindent \textbf{Part 1}.   If $p_i$ starts a new round $s$ and samples the sequence  $\langle p_{1,s},\dots p_{k,s} \rangle$ at  $t$ (line \ref{sampled}), then a calculation for the binomial shows that the probability that at most 8 (i.e.\ $k-\alpha_2$) members of this sequence are correct and locked on $\sigma_t^*$ at $t$ is $<1.18\times 10^{-20}$, i.e.\ $\text{Bin}(80,0.8\times 0.75,\leq 8)<1.18 \times 10^{-20}$.
This calculation applies to any single process beginning a single round. Next, we wish to bound the probability that the following statement holds throughout the execution: 
\begin{enumerate} 
\item[$(\dagger_0^{\diamond})$]   If a correct process starts a new round $s$ and samples the sequence  $\langle p_{1,s},\dots p_{k,s} \rangle$ at  $t$, then at least 9 members of this sequence are correct and locked on $\sigma_t^*$ at  $t$. 
\end{enumerate}
\noindent  The significance of $(\dagger_0^{\diamond})$ is essentially the same as in Section \ref{SnowFDanal}: note that if $p_i$ starts a new round $s$ and samples the sequence  $\langle p_{1,s},\dots p_{k,s} \rangle$ at  $t$, and if $p_{j,s}$ is correct and locked on $\sigma_t^*$ at $t$, then $p_{j,s}$ cannot respond to $p_i$'s request to send values for round $s$ by reporting (before $p_i$ stops collecting values for the round, after time $2\Delta$) that it has been locked on some string that is incompatible with $\sigma_t^*$ for at least $2\Delta$. Much as in Section \ref{SnowFDanal}, to bound the probability that $(\dagger_0^{\diamond})$ fails to hold, we can bound the number of processes and the number of rounds initiated by any process, and then apply the union bound. As before, we suppose that the protocol is executed for at most 1000 years by at most 10000 processes, with each process initiating at most 5 rounds per second.  It follows that each process initiates less than $1.6\times 10^{11}$ rounds. Applying the union bound to our analysis above then gives a cumulative error probability less than $(1.6\times 10^{11}) \times 10^4 \times (1.18\times 10^{-20})  < 1.9\times 10^{-5}$, meaning that $(\dagger_0^{\diamond})$ fails to hold  with probability at most  $1.9\times 10^{-5}$. 

\vspace{0.2cm} 
Extending this idea, let $P$ be the set of correct processes that are locked on $\sigma_t$ for the entirety of the interval $[t-2\Delta,t]$. We aim to show that all processes in $P$ remain locked on $\sigma_t$ thereafter (except with small probability). Towards a contradiction, suppose there is some first timeslot $t^*>t$ at which $p_i\in P$ ceases to be locked on $\sigma_t$, after receiving sufficiently many values for some round $s$ (line \ref{41}). Since $p_i$ ceases to update its local values $\mathtt{rlock}(j,s)$ time $2\Delta$ after initiating the round (see line \ref{2Deltad}), $p_i$ must initiate round $s$ in the interval $[t^*-2\Delta, t^*)$. Since all members of $P$ are locked on $\sigma_t$ at this time (by our choice of $t^*$), $(\dagger_0^{\diamond})$ together with the observation that follows its statement above give an immediate contradiction. It therefore follows from $(\dagger_0^{\diamond})$ that: 

\begin{enumerate} 
\item[$(\dagger_1^{\diamond})$] For all $t'\geq t$, $\sigma_{t'}\supseteq \sigma_t$. 
\end{enumerate}



\vspace{0.2cm} 
\noindent \textbf{Part 2}.  Now suppose that correct process $p_i$ initiates (forms its sample sequence for) round $s$ at $t$. Consider the probability, $x$ say, that there exists some $\sigma \not \subseteq \sigma_t$ such that $p_i$ samples 72 or more locked values in round $s$  extending  $\sigma$.  Calculations for the binomial distribution show that (independent of events prior $p_i$ initiating round $s$), this probability is less than 0.0131. To see this, note that $x<x_0+x_1+x_2$, where: 
\begin{itemize} 
\item $x_0$ is the probability that $p_i$ samples 72 or more locked values in round $s$  extending $\sigma_t\ast 0$. 
\item $x_1$ is the probability that $p_i$ samples 72 or more locked values in round $s$  extending $\sigma_t\ast 1$. 
\item $x_2$ is the probability that $p_i$ samples 72 or more locked values in round $s$ that are incompatible with $\sigma_t$. 
\end{itemize}  
Note that, for correct $p_j$ to report its locked value as extending $\sigma$, $p_j$ must be locked on $\sigma$ for at least $4\Delta$ (line \ref{3D}), meaning that (if $p_i$ receives the response within time $2\Delta$ of initiating the round) $p_j$ must have been locked on $\sigma$ for at least $2\Delta$ at $t$. A direct calculation for the binomial then shows that $x_2\leq \text{Bin}(80,0.2+(0.8*0.25),72)<  1.18 \times 10^{-20}$.   To bound $x_0$ and $x_1$, suppose first that at least 50\% of correct processes  have been locked on values extending $\sigma_t\ast 0$ for at least $2\Delta$ at $t$. In this case, $x_0$ is at most $\text{Bin}(80,(0.75\times 0.8)+0.2,\geq 72)<0.01309$, while $x_1$ is at most  $\text{Bin}(80,(0.5\times 0.8)+0.2,\geq 72)<3\times 10^{-9}$. If less than  50\% of correct processes have been locked on values extending $\sigma_t\ast 0$ for at least $2\Delta$ at $t$, then $x_0<3\times 10^{-9}$, while $x_1<0.01309$. Either way $x_0+x_1+x_2<0.0131$, as claimed. 
This calculation held irrespective of events prior to $p_i$ initiating round $s$.  So, if we consider any 12 given consecutive rounds $[s,s+11]$ and any given correct process $p_i$ that initiates round each round $s'\in [s,s+11]$ at $t_{s'}$ (say), the probability that, for every $s'\in [s,s+11]$,  $p_i$ samples at least 72 locked values in round $s'$ that are not extended by $\sigma_{t_{s'}}$  is upper bounded by  $0.0131^{12}<10^{-22}$. If at most 10,000 processes execute the protocol for at most 1000 years, executing at most 5 rounds per second, we can then apply the union bound to conclude that the following statement fails to hold with probability at most $10^{-22}\times 10000 \times (1.6 \times 10^{11})<  2\times 10^{-7}$:

\begin{enumerate}
\item[$(\dagger_2^{\diamond})$] If a correct process finalizes some string  $\sigma$ at $t$, then there exists $t'<t$ for which $\sigma_{t'}\supseteq \sigma$.  
\end{enumerate}

\vspace{0.2cm} 
\noindent \textbf{Part 3}.  Now we put parts 1 and 2 together. From the union bound and the analysis above, we may conclude that $(\dagger_1^{\diamond})$ and $(\dagger_2^{\diamond})$ both hold, except with probability at most 
$(1.9\times 10^{-5})+( 2\times 10^{-7})<2\times 10^{-5}$. Consistency then follows directly from $(\dagger_1^{\diamond})$ and $(\dagger_2^{\diamond})$.

\section{Producing conditions for quick finality} \label{qf} 

The previous sections give a rigorous proof of consistency for Snowman$^{\diamond}$ in the partially synchronous setting. The conditions required for finalization are quite strict, because they are required to ensure that, even in the case that a 20\% adversary attacks the protocol for 1000 years, the probability that any two correct processes will ever finalize incompatible values is small. The requirement for consistency in partial synchrony also causes some increase in latency, because, without any assumption on clock synchronization, processes are required to be locked on a string for at least $4\Delta$ (line \ref{3D} of Algorithm 3) before others can safely interpret their response as guaranteeing that they had been locked for at least $2\Delta$ at the start of the respective round (if correct). 
 In this section, we consider two approaches to reducing latency: 

 \vspace{0.2cm} 
 \noindent \textbf{Introducing some assumptions on clock synchronization}. Apple watches are reportedly synchronized to within 50 milliseconds of UTC.\footnote{See, for example, \url{https://www.theverge.com/2015/3/9/8176793/apple-watch-precision-accuracy-timekeeping}.} In an age where precise synchronization is possible, it may be reasonable to assume (even when partial synchrony applies to message delivery) that the clocks of correct processes are synchronized to within some bound $\Delta^*$. This assumption will only be useful when $\Delta^*$ is small compared to $\Delta$.

 \vspace{0.2cm} 
 \noindent \textbf{Introducing conditions for temporary finality}. The basic idea is that, while processes should use the instructions of Section \ref{SnowmanD} to determine when to extend their local value $\mathtt{final}$ (never to redefine it below that length again), they may also be willing to regard certain values as `temporarily' final in the face of less evidence. This `temporary' finalization of values has no impact on other instructions carried out by the protocol,  but may be used to achieve low latency in a scenario where one is prepared to accept a $10^{-6}$ chance, say, that a particular transaction will be reverted (and where one may be willing to assume stronger bounds on the adversary).  

\vspace{0.2cm}
In the following subsections, we expand on both of these approaches. By combining these approaches, the basic aim is that, under `standard operating conditions' in which the vast majority of processes act correctly most of the time, the process of temporary block finalization should proceed (roughly) as follows during periods of synchrony: 
\begin{enumerate} 
\item Block propagation; 
\item Each process carries out a single sampling round, in which they lock on the block; 
\item Each correct process carries out one further round of sampling, in which they find that a large proportion of processes are locked on the block, and which suffices to temporarily finalize the block.  
\end{enumerate}

\subsection{Introducing some assumptions on clock synchronization}

Suppose we know that correct processes have clocks synchronized to within $\Delta^*$. In this case, we can make a small modification to the protocol. When processes request values for a particular round, they should also report the current time $t$ on their local clock. This means that they now send a message $(\mathtt{s},t)$, rather than just the message $\mathtt{s}$ in line \ref{ask}. Then we replace lines \ref{resps}-\ref{resps2} with: 

\vspace{0.2cm} 
\noindent 51.\hspace{0.1cm} \textbf{If} $p_i$ has received $(s,t')$ from $p_j$ (for the first time):  

 \noindent 52. \hspace{0.2cm} Let $\sigma$ be the longest initial segment of $\mathtt{pref}$ s.t. $\mathtt{locktime}(\sigma)\downarrow \leq t'-2\Delta- \Delta^*$; 

\noindent  53. \hspace{0.2cm}  Send $(s,\mathtt{chain}(\mathtt{pref}),\sigma)$ to $p_j$; 

\vspace{0.2cm} 
\noindent If $\Delta^*$ and actual message delays are smaller than $\Delta$, then this will reduce latency (by up to $2\Delta$). 

\subsection{Introducing conditions for temporary finality}
\textbf{Assumptions}. We suppose clocks are synchronized to within $\Delta^*$ and that a process may be willing to temporarily finalize transactions, so long as they can be quite sure that they will not be reverted within the next hour (during which time we expect the temporarily finalised transaction to be fully finalized\footnote{Strictly speaking, this requires some assumption on liveness, which we do not address in this paper. We note, however, that the Frosty module, described in \cite{buchwald2024frosty} can be used to ensure the required liveness.}). As an example, we suppose that an arbitrary proportion of processes may be subject to crash or omission faults, but that, in the hour before and after we decide whether to `temporarily finalize' a transaction, at most 10\% of processes will be actually be Byzantine (and actively carrying out instructions that are not as prescribed by the protocol). If this assumption is incorrect, then temporarily finalized transactions may be reverted, but values finalized by the protocol as described in Section  \ref{SnowmanD} will not be (so long as the assumptions of that section hold).

\vspace{0.2cm}
\noindent \textbf{Protocol modifications}. We now replace lines  \ref{resps}-\ref{resps2} with: 

\vspace{0.2cm} 
\noindent 51.\hspace{0.1cm} \textbf{If} $p_i$ has received $(s,t')$ from $p_j$ (for the first time):  

 \noindent 52. \hspace{0.2cm} Let $\sigma$ be the longest initial segment of $\mathtt{pref}$ s.t. $\mathtt{locktime}(\sigma)\downarrow \leq t'-2\Delta- \Delta^*$; 

  \noindent 53. \hspace{0.2cm} Let $\sigma'$ be the longest initial segment of $\mathtt{pref}$ s.t. $\mathtt{locktime}(\sigma')\downarrow \leq t'- \Delta^*$ and such that $p_i$'s 
  
\hspace{0.32cm}   local $\mathtt{pref}$ value has not been incompatible with $\sigma'$  since $t'- \Delta^*- 2\Delta $;

\noindent  54. \hspace{0.2cm}  Send $(s,\mathtt{chain}(\mathtt{pref}),\sigma, \sigma')$ to $p_j$; 

\vspace{0.1cm} 
If a process $p_i$ is locked on $\sigma$ at $t$ and if  $p_i$'s local $\mathtt{pref}$ value has not been incompatible with $\sigma$  since $t- 2\Delta$, then we say that $p_i$ is \emph{safely} locked on $\sigma$ at $t$.  

\vspace{0.2cm} 
\noindent \textbf{The conditions for temporary finality}. Process $p_i$ temporarily finalizes $\sigma''$ upon finding that at least $\alpha_2$ (i.e. 72 out of 80) sample responses for round $s$ are of the form  $(s,\mathtt{chain}(\mathtt{pref}),\sigma, \sigma')$ such that $\sigma''\subseteq \sigma'$. If $p_j$ sends the response $(s,\mathtt{chain}(\mathtt{pref}),\sigma, \sigma')$ to $p_i$, we also say that $p_i$ \emph{reports being safely locked on $\sigma'$ in round $s$}. 

\vspace{0.2cm} 
\noindent \textbf{Analysis}. Let us recast the argument of Section \ref{consisanal}, replacing $\sigma_t$ and $\sigma_t^*$ with different values. We define $\tau_t$ to be the longest string $\sigma$ such that at least 60\% of non-Byzantine processes are safely locked on extensions of $\sigma$ at $t$, and we define $\tau_t^*$ to be the longest (potentialy infinite) string $\sigma$ such that at least 60\% of non-Byzantine processes have local $\mathtt{pref}$ values compatible with $\sigma$ at $t$.
  Using a similar argument to that in Section \ref{consisanal}, we will now be able to show that (given our setup assumptions)  the chance that a given temporarily finalized transaction will later be reverted is at most $10^{-6}$.  
Once again, the analysis consists of three parts.

\vspace{0.2cm} 
\noindent \textbf{Part 1}.   Suppose $p_i$ starts a new round $s$ and samples the sequence  $\langle p_{1,s},\dots p_{k,s} \rangle$ at  $t$ (line \ref{sampled}). Then a calculation for the binomial shows that the probability that at most 8 (i.e.\ $k-\alpha_2$) members of this sequence are non-Byzantine and have $\mathtt{pref}$ values compatible with $\tau_t^*$ at $t$ is $<2\times 10^{-16}$, i.e.\ $\text{Bin}(80,0.9\times 0.6,\leq 8)<2 \times 10^{-16}$.
This calculation applies to any single process beginning a single round. Next, we wish to bound the probability that the following statement holds for a given $t$: 
\begin{enumerate} 
\item[$(\clubsuit_0^{\diamond})$]   If a non-Byzantine process starts a new round $s$ and samples the sequence  $\langle p_{1,s},\dots p_{k,s} \rangle$ at  $t'\in [t-2\Delta, t+\text{ 1 hour}]$, then at least 9 members of this sequence are non-Byzantine and have $\mathtt{pref}$ values compatible  $\tau_{t'}^*$ at  $t'$. 
\end{enumerate}
\noindent   To bound the probability that $(\clubsuit_0^{\diamond})$ fails to hold for a given $t$,  we can again apply the union bound. Suppose there are at most 10000 processes, with each process initiating at most 5 rounds per second and that $\Delta$ is at most 1 minute.  This means that each process initiates less than  $2\times 10^{4}$ rounds in the interval $[t-2\Delta, t+\text{ 1 hour}]$. The union bound thus gives a cumulative error probability of less than $(2\times 10^{4}) \times 10^4 \times (2\times 10^{-16})  = 4\times 10^{-8}$, meaning that $(\clubsuit_0^{\diamond})$ fails to hold for a given $t$ with probability at most  $4\times 10^{-8}$. 

\vspace{0.2cm} 
Next, we use a similar argument as in Section \ref{consisanal}  to show that, so long as $(\clubsuit_0^{\diamond})$ holds for all $t'$ in the interval $[t-2\Delta,t+\text{ 1 hour}]$, $(\clubsuit_1^{\diamond})$ below also holds for $t$: 

\begin{enumerate} 
\item[$(\clubsuit_1^{\diamond})$] For all $t'\in [t,t+\text{ 1 hour}]$, $\tau_{t'}\supseteq \tau_t$. 
\end{enumerate}

\noindent To see this, let $P$ be the set of correct processes that are safely locked on $\tau_t$ at $t$.   We aim to show that all processes in $P$ remain safely locked on $\tau_t$ for the entirety of the interval $[t,t+\text{ 1 hour}]$.  Towards a contradiction, suppose there is some first timeslot $t^*>t$ at which $p_i\in P$ ceases to be safely locked on $\tau_t$, after receiving sufficiently many values for some round $s$ (line \ref{41}). Since $p_i$ ceases to update its local values $\mathtt{rlock}(j,s)$ time $2\Delta$ after initiating the round (see line \ref{2Deltad}), $p_i$ must initiate round $s$ in the interval $[t^*-2\Delta, t^*)$. Since all members of $P$ have local $\mathtt{pref}$ values compatible with $\tau_t$ at this time, meaning that $\tau_{t^*}^*$ extends $\tau_t$,  $(\clubsuit_0^{\diamond})$  gives an immediate contradiction.



\vspace{0.2cm} 
\noindent \textbf{Part 2}.  Now suppose that process $p_i$ initiates round $s$ at $t$. Consider the probability, $x$ say, that there exists some $\sigma \not \subseteq \tau_t$ such that $p_i$ samples 72 or more safely locked values in round $s$  extending  $\sigma$.  Calculations for the binomial distribution show that (independent of events prior to $p_i$ initiating round $s$), this probability is less than $3\times 10^{-7}$. To see this, note that $x<x_0+x_1+x_2$, where: 
\begin{itemize} 
\item $x_0$ is the probability that $p_i$ samples 72 or more safely locked values in round $s$  extending $\tau_t\ast 0$. 
\item $x_1$ is the probability that $p_i$ samples 72 or more safely locked values in round $s$  extending $\tau_t\ast 1$. 
\item $x_2$ is the probability that $p_i$ samples 72 or more safely locked values in round $s$ that are incompatible with $\tau_t$. 
\end{itemize}  
 A direct calculation for the binomial shows that $x_2\leq \text{Bin}(80,0.1+(0.9*0.4),\geq 72)<  2 \times 10^{-16}$.   To bound $x_0$ and $x_1$, suppose first that at least 50\% of non-Byzantine processes  are safely locked on values extending $\tau_t\ast 0$ at $t$. In this case, $x_0$ is at most $\text{Bin}(80,(0.6\times 0.9)+0.1,\geq 72)<2\times 10^{-7}$, while $x_1$ is at most  $\text{Bin}(80,(0.5\times 0.9)+0.1,\geq 72)<2\times 10^{-11}$. If less than  50\% of correct processes are safely  locked on values extending $\tau_t\ast 0$  at $t$, then $x_0<2\times 10^{-11}$, while $x_1<2\times 10^{-7}$. Either way $x_0+x_1+x_2<3\times 10^{-7}$, as claimed. 
The following statement therefore fails to hold with probability at most $3\times 10^{-7}$ for a given $p_i$ and $t$:  
\begin{enumerate}
\item[$(\clubsuit_2^{\diamond})$] If non-Byzantine $p_i$ samples 72 or more safely locked values extending $\sigma$ in a round initiated at $t$, then $\sigma \subseteq \tau_t$.  
\end{enumerate}

\vspace{0.2cm} 
\noindent \textbf{Part 3}. Now we put the previous observations together. If non-Byzantine $p_i$ samples 72 or more safely locked values extending $\sigma$ in a round initiated at $t$, then we have observed that, except with probability at most $3\times 10^{-7}$, $\sigma \subseteq \tau_t$. We have also deduced that $(\clubsuit_1^{\diamond})$ holds for $t$, except with probability at most $8\times 10^{-8}$. Recall the definition of $\sigma_t$ and $(\dagger_2^{\diamond})$ from Section \ref{consisanal}, and note that $\sigma_t\subseteq \tau_t$. It was shown in that section that (even under the assumption of a 20\% adversary and with an arbitrary number of crash or omission failures) $(\dagger_2^{\diamond})$ holds throughout the execution (and for all non-Byzantine processes), except with probability at most $2\times 10^{-7}$. It follows that, except with probability at most $6\times 10^{-7}$, no non-Byzantine process can finalize a value incompatible with $\sigma$ during the interval $[t,t+\text{ 1 hour}]$.

\section{Related Work} \label{rw}
There is an extensive literature that considers a closely related family of models, from \emph{voter models} \cite{holley1975ergodic} as studied in applied probability and other fields, to the \emph{Ising model} \cite{brush1967history} as studied in statistical mechanics, to the \emph{Schelling model of segregation} \cite{schelling1969models} as studied by economists and more recently by computer scientists \cite{brandt2012analysis,barmpalias2014digital} and physicists \cite{omidvar2017self,omidvar2021improved,ortega2021schelling}. This family of models has many variants, but a standard approach is to consider a process  proceeding in rounds. In each round, each participant samples some number of other participants to learn their present state, and then updates their own state according to given rule. A basic difference with our analysis is that, with two exceptions (mentioned below), these models do not incorporate the possibility of Byzantine action. Examples of research along these lines aimed specifically at the task of reaching consensus include \cite{becchetti2016stabilizing,doerr2011stabilizing,elsasser2017brief,ghaffari2018nearly,cooper2014power,cruciani2021phase} (see \cite{becchetti2020consensus} for an overview). Amongst these papers, we are only aware of \cite{becchetti2016stabilizing} and \cite{doerr2011stabilizing} considering Byzantine action. Those two papers deal only with an $O(\sqrt{n})$ adversary. 

\vspace{0.2cm} 
As noted previously, the Snow family of consensus protocol was introduced in \cite{rocket2019scalable}. Subsequent to that paper, Amores-Sesar, Cachin and Tedeschi \cite{amores2022spring} gave a complete description of the Avalanche protocol\footnote{The Avalanche protocol is a DAG-based variant of Snowman that does not aim to produce a total ordering on transactions, and was only described at a high level in \cite{rocket2019scalable}. It is not used in the present instantiation of the Avalanche blockchain.} and formally established security properties for that protocol, given an $O(\sqrt{n})$ adversary and assuming that the Snowball protocol (a variant of Snowflake) solves probabilistic Byzantine Agreement for such adversaries. The authors also described, and provided a solution for, a liveness attack. As noted in \cite{amores2022spring}, the original implementation of the Avalanche protocol used by the Avalanche blockchain (before replacing Avalanche with a version of Snowman) had already introduced modifications avoiding the possibility of such attacks. 

\vspace{0.2cm} 
In \cite{amores2024analysis}, Amores-Sear, Cachin and Schneider consider the Slush protocol, which is a member of the Snow family designed for solving Binary Byzantine Agreement. They show that coming close to a consensus already requires a minimum of $\Omega \left( \frac{\text{log }n}{\text{log }k} \right)$ rounds,
even in the absence of adversarial influence. That paper also shows that Slush reaches a stable consensus in $O(\text{log }n)$ rounds, and that this holds even when the adversary can influence up to $O(\sqrt{n})$ processes. The $\Omega \left( \frac{\text{log }n}{\text{log }k} \right)$ lower bound holds for Snowflake and Snowball.

\vspace{0.2cm} 
FPC-BI \cite{popov2021fpc,popov2022voting} is a protocol which is closely related to the Snow family of consensus protocols, but which takes a different approach to the liveness issue (for adversaries which are larger than $O(\sqrt{n})$) than that described here. The basic idea behind their approach is to use a common random coin to dynamically and unpredictably set threshold parameters (akin to $\alpha_1$ and $\alpha_2$ here) for each round, making it much more difficult for an adversary to keep the honest population split on their preferred values. Since the use of a common random coin involves practical trade-offs, their approach and ours may be seen as complementary.


\begin{thebibliography}{10}

\bibitem{amores2024analysis}
Ignacio Amores-Sesar, Christian Cachin, and Philipp Schneider.
\newblock An analysis of avalanche consensus.
\newblock {\em arXiv preprint arXiv:2401.02811}, 2024.

\bibitem{amores2022spring}
Ignacio Amores-Sesar, Christian Cachin, and Enrico Tedeschi.
\newblock When is spring coming? a security analysis of avalanche consensus.
\newblock {\em arXiv preprint arXiv:2210.03423}, 2022.

\bibitem{bagaria2019prism}
Vivek Bagaria, Sreeram Kannan, David Tse, Giulia Fanti, and Pramod Viswanath.
\newblock Prism: Deconstructing the blockchain to approach physical limits.
\newblock In {\em Proceedings of the 2019 ACM SIGSAC Conference on Computer and
  Communications Security}, pages 585--602, 2019.

\bibitem{barmpalias2014digital}
George Barmpalias, Richard Elwes, and Andy Lewis-Pye.
\newblock Digital morphogenesis via schelling segregation.
\newblock In {\em 2014 IEEE 55th Annual Symposium on Foundations of Computer
  Science}, pages 156--165. IEEE, 2014.

\bibitem{becchetti2020consensus}
Luca Becchetti, Andrea Clementi, and Emanuele Natale.
\newblock Consensus dynamics: An overview.
\newblock {\em ACM SIGACT News}, 51(1):58--104, 2020.

\bibitem{becchetti2016stabilizing}
Luca Becchetti, Andrea Clementi, Emanuele Natale, Francesco Pasquale, and Luca
  Trevisan.
\newblock Stabilizing consensus with many opinions.
\newblock In {\em Proceedings of the twenty-seventh annual ACM-SIAM symposium
  on Discrete algorithms}, pages 620--635. SIAM, 2016.

\bibitem{brandt2012analysis}
Christina Brandt, Nicole Immorlica, Gautam Kamath, and Robert Kleinberg.
\newblock An analysis of one-dimensional schelling segregation.
\newblock In {\em Proceedings of the forty-fourth annual ACM symposium on
  Theory of computing}, pages 789--804, 2012.

\bibitem{brush1967history}
Stephen~G Brush.
\newblock History of the lenz-ising model.
\newblock {\em Reviews of modern physics}, 39(4):883, 1967.

\bibitem{buchwald2024frosty}
Aaron Buchwald, Stephen Buttolph, Andrew Lewis-Pye, Patrick O'Grady, and Kevin
  Sekniqi.
\newblock Frosty: Bringing strong liveness guarantees to the snow family of
  consensus protocols.
\newblock {\em arXiv preprint arXiv:2404.14250}, 2024.

\bibitem{cooper2014power}
Colin Cooper, Robert Els{\"a}sser, and Tomasz Radzik.
\newblock The power of two choices in distributed voting.
\newblock In {\em International Colloquium on Automata, Languages, and
  Programming}, pages 435--446. Springer, 2014.

\bibitem{cruciani2021phase}
Emilio Cruciani, Hlafo~Alfie Mimun, Matteo Quattropani, and Sara Rizzo.
\newblock Phase transitions of the k-majority dynamics in a biased
  communication model.
\newblock In {\em Proceedings of the 22nd International Conference on
  Distributed Computing and Networking}, pages 146--155, 2021.

\bibitem{doerr2011stabilizing}
Benjamin Doerr, Leslie~Ann Goldberg, Lorenz Minder, Thomas Sauerwald, and
  Christian Scheideler.
\newblock Stabilizing consensus with the power of two choices.
\newblock In {\em Proceedings of the twenty-third annual ACM symposium on
  Parallelism in algorithms and architectures}, pages 149--158, 2011.

\bibitem{dolev1985bounds}
Danny Dolev and R{\"u}diger Reischuk.
\newblock Bounds on information exchange for byzantine agreement.
\newblock {\em Journal of the ACM (JACM)}, 32(1):191--204, 1985.

\bibitem{elsasser2017brief}
Robert Els{\"a}sser, Tom Friedetzky, Dominik Kaaser, Frederik Mallmann-Trenn,
  and Horst Trinker.
\newblock Brief announcement: rapid asynchronous plurality consensus.
\newblock In {\em Proceedings of the ACM symposium on principles of distributed
  computing}, pages 363--365, 2017.

\bibitem{ghaffari2018nearly}
Mohsen Ghaffari and Johannes Lengler.
\newblock Nearly-tight analysis for 2-choice and 3-majority consensus dynamics.
\newblock In {\em Proceedings of the 2018 ACM Symposium on Principles of
  Distributed Computing}, pages 305--313, 2018.

\bibitem{holley1975ergodic}
Richard~A Holley and Thomas~M Liggett.
\newblock Ergodic theorems for weakly interacting infinite systems and the
  voter model.
\newblock {\em The annals of probability}, pages 643--663, 1975.

\bibitem{lewis2024lumiere}
Andrew Lewis-Pye, Dahlia Malkhi, Oded Naor, and Kartik Nayak.
\newblock Lumiere: Making optimal bft for partial synchrony practical.
\newblock In {\em Proceedings of the 43rd ACM Symposium on Principles of
  Distributed Computing}, pages 135--144, 2024.

\bibitem{lewis2023permissionless}
Andrew Lewis-Pye and Tim Roughgarden.
\newblock Permissionless consensus.
\newblock {\em arXiv preprint arXiv:2304.14701}, 2023.

\bibitem{naor2024expected}
Oded Naor and Idit Keidar.
\newblock Expected linear round synchronization: The missing link for linear
  byzantine smr.
\newblock {\em Distributed Computing}, 37(1):19--33, 2024.

\bibitem{plusplus}
Patrick O'Grady.
\newblock Apricot phase four: Snowman++ and reduced c-chain transaction fees.
\newblock {\em
  \url{https://medium.com/avalancheavax/apricot-phase-four-snowman-and-reduced-c-chain-transaction-fees-1e1f67b42ecf}}.

\bibitem{omidvar2017self}
Hamed Omidvar and Massimo Franceschetti.
\newblock Self-organized segregation on the grid.
\newblock In {\em Proceedings of the ACM Symposium on Principles of Distributed
  Computing}, pages 401--410, 2017.

\bibitem{omidvar2021improved}
Hamed Omidvar and Massimo Franceschetti.
\newblock Improved intolerance intervals and size bounds for a schelling-type
  spin system.
\newblock {\em Journal of Statistical Mechanics: Theory and Experiment},
  2021(7):073302, 2021.

\bibitem{ortega2021schelling}
Diego Ortega, Javier Rodr{\'\i}guez-Laguna, and Elka Korutcheva.
\newblock A schelling model with a variable threshold in a closed city
  segregation model. analysis of the universality classes.
\newblock {\em Physica A: Statistical Mechanics and its Applications},
  574:126010, 2021.

\bibitem{popov2021fpc}
Serguei Popov and William~J Buchanan.
\newblock Fpc-bi: Fast probabilistic consensus within byzantine
  infrastructures.
\newblock {\em Journal of Parallel and Distributed Computing}, 147:77--86,
  2021.

\bibitem{popov2022voting}
Serguei Popov and Sebastian M{\"u}ller.
\newblock Voting-based probabilistic consensuses and their applications in
  distributed ledgers.
\newblock {\em Annals of Telecommunications}, pages 1--23, 2022.

\bibitem{rocket2019scalable}
Team Rocket, Maofan Yin, Kevin Sekniqi, Robbert van Renesse, and Emin~G{\"u}n
  Sirer.
\newblock Scalable and probabilistic leaderless bft consensus through
  metastability.
\newblock {\em arXiv preprint arXiv:1906.08936}, 2019.

\bibitem{schelling1969models}
Thomas~C Schelling.
\newblock Models of segregation.
\newblock {\em The American economic review}, 59(2):488--493, 1969.

\bibitem{shrestha2024sailfish}
Nibesh Shrestha, Rohan Shrothrium, Aniket Kate, and Kartik Nayak.
\newblock Sailfish: Towards improving latency of dag-based bft.
\newblock {\em Cryptology ePrint Archive}, 2024.

\bibitem{spiegelman2022bullshark}
Alexander Spiegelman, Neil Giridharan, Alberto Sonnino, and Lefteris
  Kokoris-Kogias.
\newblock Bullshark: Dag bft protocols made practical.
\newblock In {\em Proceedings of the 2022 ACM SIGSAC Conference on Computer and
  Communications Security}, pages 2705--2718, 2022.

\bibitem{yin2019hotstuff}
Maofan Yin, Dahlia Malkhi, Michael~K Reiter, Guy~Golan Gueta, and Ittai
  Abraham.
\newblock Hotstuff: Bft consensus with linearity and responsiveness.
\newblock In {\em Proceedings of the 2019 ACM Symposium on Principles of
  Distributed Computing}, pages 347--356, 2019.

\end{thebibliography}
\end{document}